\documentclass[aps]{revtex4}
\usepackage{amssymb}
\usepackage{graphicx}
\begin{document}

\title{\small Stability of the synchronization manifold in nearest neighbors non identical van der Pol-like oscillators}

\author{R. Yamapi}
\affiliation{ Fundamental Physic's
Laboratory, Group of Nonlinear Physics and Complex System,
Department of Physics, Faculty of Science, University of Douala,
Box 24 157 Douala,
Cameroon, 
Email address: ryamapi@yahoo.fr}

\author{H. G. Enjieu Kadji}
\affiliation{Institute for Development, Aging and Cancer, Tohoku
University 4-1 Seiryocho, Aobaku, Sendai 980-8575, Japan, 
Email address: henjieu@yahoo.com}

\author{G. Filatrella}
\affiliation{Laboratorio Regionale SuperMat, CNR/INFM Salerno and
Dipartimento di Scienze Biologiche ed Ambientali, Universit\`a del
Sannio, via Port'Arsa 11, I-82100 Benevento, Italy, 
Email address: giofil@sa.infn.it}

\date{January 19, 2010 }

\begin{abstract}

We investigate the stability of the synchronization manifold in a
ring and an open-ended chain of nearest neighbors coupled
self-sustained systems, each self-sustained system consisting of
multi-limit cycles van der Pol oscillators. Such model represents,
for instance, coherent oscillations in biological systems through
the case of an enzymatic-substrate reaction with ferroelectric
behavior in brain waves model. The ring and open-ended chain of
identical and non-identical oscillators are considered separately.
By using the Master Stability Function approach (for the identical
case) and the complex Kuramoto order parameter (for the
non-identical case), we derive the stability boundaries of the
synchronized manifold. We have found that synchronization occurs
in a system  of many coupled modified van der Pol oscillators and
it is stable even in presence of a spread of parameters.

\end{abstract}
 \maketitle

\section{Introduction }
\noindent

During the last decades, the emergence of collective dynamics in
large networks of coupled units has been investigated in
disciplines such as physics, chemistry, biology and ecology. In
particular, the effect of synchronization in systems of coupled
oscillators nowadays provides a unifying framework for different
phenomena observed in nature (for some reviews see
\cite{Pikovsky-Rosemblum-Kurths-2001,boccaletti-kurth-2002,
Manrubia-Mikhailov-Zanette-2004}). Recently, complex networks have
provided a challenging framework for the study of synchronization
of dynamical units, based on the interplay between complexity in
the overall topology and local dynamical properties of the coupled
units.  A key problem is to assess conditions that guarantee the
stability of the synchronous behavior for a network topology with
some class of coupling configuration. To determine such
conditions, many methods have been developed  and among these, the
so-called Master Stability Function (MSF)
\cite{pecora-carroll-1998}. Indeed, the MSF approach was
originally introduced for arrays of coupled oscillators
\cite{pecora-carroll-1998}, and it has been later extended to the
case of a complex networks of dynamical systems coupled with
arbitrary topologies
\cite{Barahona-Pecora-2002,Hu-Yang-Liu-1998,Zhan-Hu-Yang-2000,
Chen-Rangarajan-Ding-2003}. We will use the MSF to investigate the
stability boundaries of the synchronization manifold in a ring and
an open-ended chain of self-sustained oscillators described by
coupled multi-limit cycles van der Pol oscillators (MLC-vdPo) that
might display birhythmicity,
 a mathematical model for instance suitable for
 some enzymatic substrate reaction with
ferroelectric behavior in a brain waves model \cite{kaiser1}, as
well as in other biochemical models \cite{decroly-goldbeter}.

Limit cycles are the fascinating objects in natural sciences where
nonlinear kinetics due to feedback and co-operativity leads to
instability of the fix point  and evolves towards periodic
sustained oscillations. Examples are abundant, with periods
ranging from cardiac rhythms of seconds, glycolysis over minutes,
circadian oscillations over the 24 hours, while epidemiological
oscillations extend over the years
\cite{goldbeter-1996,winfree,goldbeter-2002}.
Compared to the classical van der Pol system that
possesses only one stable limit cycle, the MLC-vdPos has the
advantage to  better explain some biological processes, through
its possibility of showing up  multi limit cycles oscillatory
states \cite{kaiserv}. Thereby, we aim to understand how a large
(but finite) number of interacting MLC-vdPos will behave
collectively, given their individual dynamics and a corresponding
coupling topology. We will investigate two oscillators coupling
topology, a ring and an open-ended chain, both with nearest
neighbors coupling. Indeed, such couplings are of paramount
importance since they occur in many applications, both in natural
and artificial systems, that involve  the problem of coordination
of multiple agents (circadian rhythm, contraction of coronary
pacemaker cells, firing of memory neurons in the brain,
superconducting Josephson junction arrays, design of oscillator
circuits, sensor
networks)\cite{jeong-chi,schenato-songwai-sastry-bose,richard-ivry-thomas,michael,wiesenfeld-colet-strogatz,paolaPRL}.
We will show that synchronization also occurs in the system of many
coupled modified van der Pol oscillators and it is stable even in
presence of a spread of parameters.

The organization of the paper is the following: Section II deals
with the description of the model under consideration. In Section
III, the stability boundaries of the synchronized states in a ring
of identical MLC-vdPos are retrieved through the MSF  approach, and
numerical simulations are used to validate and complement the
results. Section IV addresses the ring of non-identical MLC-vdPos,
while Section V considers the stability of synchronization
manifold in a chain of open-ended nearest neighbors coupled
identical and non-identical self-sustained oscillators.
Finally, Section VI is devoted to
the conclusions.

\section{The self-sustained modified van der Pol oscillator}
\noindent

The modified van der Pol oscillator is described by the
 following nonlinear differential equation
 (in non-dimensional form)
\begin{eqnarray}
\label{eq1}
\ddot x_c-\mu (1- x_c^2+\alpha x_c^4-\beta x_c^6)\dot
x_c+x_c=0.
\end{eqnarray}
(Overdots stand for the derivative with respect to time). Such
model, that exhibits an extremely rich bifurcation behavior, was
proposed by Kaiser \cite{kaiserv}, and describes many dynamical
systems, ranging from physics, biochemistry to engineering. When
employed to model biochemical systems, namely  the
enzymatic-substrate reaction,  $\dot x_c$  represents the rate of
change of the number of excited enzyme molecules and  $x_c$ in Eq.
(1) is proportional to the population of enzyme molecules in the
excited polar state. The quantities $\alpha$ and $\beta$ are
positive parameters which measure the degree of tendency of the
system to a ferroelectric instability compared to its electric
resistance, while $\mu$ is the parameter that tunes  nonlinearity
\cite{enjieu-chabi-yamapi-woafo}. The nonlinear dynamics and the
synchronization process of two such systems have been investigated
recently \cite{enjieu-chabi-yamapi-woafo,enjieu-yamapi-chabi},
while Ref. \cite{yamapi-nana-enjieu-2007} considers  its dynamics
and active control to find that chaos can be tamed for an
appropriate choice of the coupling parameters. Coupled van der Pol
oscillators have been considered in
\cite{nana06,kuznetov09,barron09}. Depending to the values of the
parameters $\beta$ and $\alpha$, Eq.~(\ref{eq1}) can lead to one
or three limit cycles. When three limit cycles are obtained, two
of them are stable and one is unstable, a condition for
birhythmicity \cite{decroly-goldbeter}: The unstable limit cycle
represents the separatrix between the basins of attraction of the
two stable limit cycles (see \cite{enjieu-chabi-yamapi-woafo}). We
will avoid such region for simplicity, so the uncoupled elements
are monorhythmical.  We show in Fig.1 the region of existence of
birhythmicity in the two parameter phase space ($\beta$-$\alpha$)
\cite{enjieu-chabi-yamapi-woafo,enjieu-yamapi-chabi}.

\section{The ring of identical self-sustained oscillators}
\subsection{The model}
\noindent

When the coupling is realized through a ring topology, {\it i.e.}
with periodic boundary conditions, the model is described by the
following set of dimensionless nonlinear differential equations:
\begin{eqnarray}
\label{eq2}
&& \ddot x_1-\mu (1- x^2_1+\alpha
x_1^4-\beta x_1^6)\dot x_1+x_1=K(x_2-
2x_1+x_N),\nonumber\\
&& \ddot x_{\nu}-\mu (1- x^2_{\nu}+\alpha
x_{\nu}^4-\beta x_{\nu}^6)\dot x_{\nu}+x_{\nu}=K( x_{\nu+1}-
2x_{\nu}+x_{\nu-1}),\quad \nu=2,3,...,N-1,\nonumber\\
&& \ddot x_N-\mu (1- x^2_N+\alpha x_N^4-\beta x_N^6)\dot
x_N+x_N=K( x_1- 2x_N+x_{N-1}).
\end{eqnarray}
Here, $K$ stands for the diffusive (nearest-neighbor) coupling
strength. The physical meaning of the variables depends on the
nature of the system described by such a ring. For instance, let
us consider the case of a ring of immobilized self-sustained
enzymes that minimize the cost of production by permitting
repeated use of the enzymes and substantially increases the
stability of the enzyme reactions themselves
\cite{kamoun-lavoine-verneuil}. We propose to describe the
coupling between cells by means of the diffusion of the solute
concentration, so $K$ is proportional to the difference in the
concentration. Since within each self-sustained system there are
promoters (positive direction of solute flow) and inhibitors
(negative direction of solute flow) that determine the direction
of the solute flow, it follows that the coupling coefficient $K$
can be either negative or positive.

The system of  diffusive  coupling occurs in many complex systems.
For instance, the ring of multi-limit cycles oscillators is of
interest to construct hypotheses on the in vivo enzymes behavior
when they possess more than one stable limit cycle. A ring of the MLC-vdPos can therefore
serve as a simple model of a biological oscillator. The set of equations (\ref{eq2}) can also be
exploited in electronics engineering as a network of parallel
microwaves oscillators \cite{fukui-nogi-1980,fukui-nogi-1986}:
Multiple limit cycles can be obtained from a classical van der Pol
oscillator with hysteresis in the inductance \cite{appelbe}. Since
we are interested in the synchronization manifold, the following
subsection deals with the stability of the synchronous states in
the ring of mutually coupled identical self-sustained oscillators.

\subsection{Stability of the synchronization manifold}
 \noindent

We aim now to determine the stability of the synchronous state in
the system (2) of self-sustained oscillator requiring that each
of the perturbed trajectories returns to its original limit cycle.
We are therefore interested in the bifurcations
from the state that resides on a
synchronization manifold denoted by
$\mathcal{M}=$$\{(x_1,y_1)=(x_2,y_2)=...=(x_N,y_N) \}$. In order
to rewrite Eqs. (\ref{eq2}) as set of flows, we have introduced a new
variable $y_\nu=\dot x_\nu$; with such new variables the corresponding dynamic equations read:
\begin{eqnarray}
\label{eq3}
\dot x_\nu&=&y_\nu, \nonumber\\
 \dot y_{\nu}&=&\mu (1- x^2_{\nu}+\alpha x_{\nu}^4-\beta x_{\nu}^6)
y_{\nu}-x_{\nu}+K( x_{\nu+1}- 2x_{\nu}+x_{\nu-1}), \qquad
 \nu=1,2,...,N.
\end{eqnarray}
To verify the stability of the synchronization manifold
$\mathcal{M}$ we make use of the MSF approach
\cite{boccaletti-2006,pecora-carroll-1998}. Thereby, let
$\mathbf{X}^i$ be the two-dimensional vector of the dynamical
variables of the $i^{th}$ unit, $\mathbf{H}:R^2\to R^2$ an
arbitrary function describing the coupling between each unit
variables. Thus, the dynamics of the  $i^{th}$ unit is rewritten
as a function of the $2\times N$ column vector state
$\mathbf{X}^i$ as
\begin{eqnarray}
\label{eq4} \mathbf{\dot
X}^i=\mathbf{F(X^i)}+K\sum_j^NG_{ij}\mathbf{H(X^j)}, \qquad
i=1,2,...,N,
\end{eqnarray}
where $\mathbf{X}^i$=$[x_i,y_i]^T,$ $\mathbf{F(X^i)}$$=[y_i,\mu
(1- x^2_{i}+\alpha x_{i}^4-\beta x_{i}^6) y_{i}-x_{i}]^T$, and the
function $\mathbf{H}$ is defined through the matrix

\begin{displaymath}
\mathbf{ E }=\left (
\begin{array}{cc}
0 &0\\
1 &0
\end{array}
\right ),
\end{displaymath}
by $\mathbf{H(X^i)=E X^i}$. $G_{ij}\in R$ are the elements of the
$N\times N$ symmetry connectivity matrix $\mathbf{G}$ defined as
\begin{displaymath}
\mathbf{ G }=\left (
\begin{array}{ccccc}
-2&1& 0& \ldots &1\\
1 &-2&1& \ldots &0\\
0&1&-2&\ldots&0\\
\vdots &\vdots&\vdots&\ddots&\vdots\\
1&0&\ldots &1&-2
\end{array}
\right ).
\end{displaymath}

Since  the ring is made of identical self-sustained oscillators,
the evolution function $F\mathbf{(X_i)}$ in Eq. (4) is the same
for all ring node. This  ensures the existence of an invariant set
$(x_i (t),y_i(t)) = (x_s (t ),y_s(t))$, $\forall i$, representing
the complete synchronization manifold $\mathcal{M}$. Following the
MSF scheme
\cite{Pikovsky-Rosemblum-Kurths-2001,boccaletti-kurth-2002,pecora-carroll-1998},
the stability of the resulting dynamical states can be determined
by letting
\begin{eqnarray*}
&&x_\nu(t)=\delta x_\nu(t)+x_s(t),\\
&&y_\nu(t)=\delta y_\nu(t)+y_s(t),
\end{eqnarray*}
and linearizing equations (\ref{eq4}) around the periodic
limit-cycle state $(x_s,y_s)$. This approach leads to the
following equation:

\begin{eqnarray}
\label{eq5}
 \delta\mathbf{\dot X}(t)=[1_N \bigotimes
\mathbf{JF(X_s)}+K\mathbf{G} \bigotimes\mathbf{JH(X_s)}]\delta
X(t),
\end{eqnarray}
where $\bigotimes$ stands for the direct product between matrices.
$J$ denotes the Jacobian operator and the $2\times N$ column
vector $\delta \mathbf{ X}^i(t)$$=(\delta x_i(t),\delta y_i(t))$
is the deviation of the $i^{th}$ vector state from the
synchronization manifold. We have used the definitions $\mathbf{
H(X^i)= E X^i}$,
and $J\mathbf{H}=E$.\\
A necessary condition for stability of the synchronization
manifold  is that the set of $(N-1)\times 2$ Lyapunov exponents
that corresponds to phase space directions transverse to the
2-dimensional hyperplane $ \mathbf{ X}^i(t)= \mathbf{ X}_s(t)$
should be entirely made of negative values
\cite{boccaletti-2006,pecora-carroll-1998}.  The arbitrary state
$\delta \mathbf{ X}(t)$ can be written as $\delta \mathbf{
X}(t)=\sum_{ i=1}^Nv_i \bigotimes \xi_i(t)$ with $\xi_i(t)=(\xi_{
i,1}(t),\xi_{ i,2}(t))$. If one applies $[v_j]^T$ ($\gamma_i$ and
$[v_i]$ are the set of real eigenvalues and the associated
orthonormal eigenvectors of the matrix $\mathbf{G}$) to the left
side of each equation in (\ref{eq5}), one finally obtains the
following set of N variational equations

\begin{eqnarray}
\label{eq6} \mathbf{\dot \xi_k}(t)=[ \mathbf{JF(X_s)}+K
\gamma_k\mathbf{JH(X_s)}]\xi_k(t), \qquad k=0,1,2,...,N-1.
\end{eqnarray}
 The eigenvalues $\gamma_k$ of $\mathbf{G}$ are given
by $\gamma_k=-4\sin^2(\pi k/N)$
\cite{pecora-carroll-1998,pecora-1998}. Let us remark that the mode
 $k=0$ is the "slower" or uniform mode, while $k=N-1$ is the "faster" or
 more rapidly oscillating (in the momentum space) mode. Each mode $k$ in Eq.
(\ref{eq6}) corresponds to a set of 2 conditional Lyapunov
exponent $\lambda_k^j$ ($j=1,2$) along the eigenmode related  to a
specific eigenvalue $\gamma_k$. Eqs. (\ref{eq6}) enable to compute
the maximum Lyapunov exponent $\lambda_k^{max}$ of each mode $k$
as a function of the coupling parameter $K$. In the following, we
will use the parametrical behavior of the largest of such
exponents $\Lambda (K)$, called MSF, to determine the stability
boundaries of the synchronization states in the ring of coupled
self-sustained systems.

\subsection{Numerical simulations }

We  now address the numerical computation of the Lyapunov
exponents and then the MSF $\Lambda$, by solving the equations of
motion (1) and the variational equations (6), with a fourth-order
{\it Runge-Kutta} algorithm.
The parameters used are representative of the monorhythmical
 solutions: $\mu=0.1$, and in (i): $\alpha=\beta=0.1$, while
 in case (ii) we set $\alpha=0.114,\beta=0.005$.

When the coupling coefficient is turned off ($i.e.$ $K=0$), the
self-sustained systems are uncoupled and the corresponding MSF is
$\Lambda(K=0) = 0$. For $K\neq 0$, and for a given ring size $N$,
the MSF $\Lambda(K)$ will enable us to derive the range of the
coupling parameter $K$ in which the transverse {\it Fourier} modes
are stable, and therefore each oscillator of the ring exhibits the
same dynamical states. The behavior of the MSF as a function of
the coupling $K$ and the number of oscillators $N$ is plotted in
Fig. 2. Starting from large negative $K$ values, the first domain
is the unstable synchronization domain ($D_{US}$) in which there
is some positive transverse Lyapunov exponent. The boundaries
slowly change with the number $N$ of oscillators (see Table 1);
when synchronization is not observed, all the modes are on the
transverse manifold where variations transverse to the
synchronization manifold
 do not decay; all or some of the transverse {\it Lyapunov}
exponents are positive, {\it i.e.} $\lambda^{max}_k>0$.

{\scriptsize
\begin{eqnarray*}
\begin{tabular}{|c|c|c|c|}
\hline $N$&Domains of coefficient K for unstable synchronization ($D_{US}$)&  N &Domains of coefficient K for unstable synchronization ($D_{US}$) \\
\hline  4&  ]-$\infty$;-0.25[$\cup$]-0.02;0[$\cup$ ]0;+0.005 [&  60&]-$\infty$;1.93 [\\
\hline  10& ]-$\infty$;-0.25[$\cup$]-0.028;0[$\cup$]0;0.055[&  70&]-$\infty$;2.63 [\\
\hline  20& ]-$\infty$;-0.25[$\cup$]-0.11;0[$\cup$]0.0;0.21[ &  80&]-$\infty$;3.0 [\\
\hline  30& ]-$\infty$;0.48[&  90&]-$\infty$;4.36 [\\
\hline 40& ]-$\infty$;0.86 [&    100 &]-$\infty$;5.38 [\\
\hline 50 & ]-$\infty$;1.34[&   &   \\
\hline
\end{tabular}
\end{eqnarray*}
\begin{center}
{\it Table 1: Range of coupling values for unstable synchronous
state in the ring of oscillators with one limit-cycle. Parameters
of the system are $\mu=0.1$, $\alpha=0.1$, $\beta=0.1$.}
\end{center}}

In other words, the {\it Fourier} modes increase continuously or
possess a bounded oscillatory behavior and one never observes
stable synchronization in the ring for such a choice of the
coupling strength. It should be underlined that the region
$D_{US}$ where unstable synchronization is found, can be divided
in two sub-domains: the first sub-domain corresponds to the
non-synchronization phenomena, while the second sub-domain
corresponds to partial synchronization. For example for $N=4$, in the
first sub-domain ($K\in ]-\infty; -0.25[$), any perturbed
trajectory leads the oscillators to continuously drift away from
their original limit cycles. In the second sub-domain, ($K\in ]-0.02; 0 [$)
 although perfect synchronization is not observed, there is some tendency
 towards co-operative behavior, as will be shown in Sect. IV
  by means of the Kuramoto order parameter.

It appears  that as the coupling coefficient $K$ increases from
$-\infty$, where the ring of coupled self-sustained systems is on
the unstable synchronized states (and therefore all Lyapunov
exponents are positive), the mode $k=1$ is the first one to move
from the unstable to the stable domain. On the other hand, $k=N/2$
is the last mode that leaves from the unstable domain to enter
the stable state. At this point, the ring is on the stable
synchronized state; when the coupling coefficient $K$ is further
increased, the mode $k=1$ is the first mode to move from the
stable domain to the unstable one, and the mode $k=N/2$ is the
last one. For the case of the positive  coupling coefficient $K$,
the reverse situation is observed: the $k=N/2$ mode is the first
to become stable. Also in the case (ii) with $\alpha=0.114$ and
$\beta=0.005$, is found a similar behavior.

Since from a biological point of view, negative and positive
values of the coupling strength refer to inhibitory and excitatory
process, respectively, one can conclude the following: The
instability of a synchronization process within a system of
inhibitory ($K<0$) cells is driven by the slowest, or lower $k$,
mode. On the other hand, when a system is made of excitatory
($K>0$) cells, the instability  of the process is initiated
through the fastest, or higher $k$, mode.

Fig. 3 shows the stability diagram in the $(N,K)$ plane, {\it i.e.
} it displays the border between the two main dynamical states. We
have found with the MSF method that the number of units strongly
affects the stability boundaries of the synchronization process in
the ring. In particular, when $N>30$, there is no domain of stable
synchronization for negative values of the coupling coefficient
$K$.
It is important to notice that Eqs. (\ref{eq6}) have been linearized around the stable orbit.
In order to check the validity of the above semi-analytical
results, we have numerically solved Eqs. (\ref{eq2}). Numerically,
the stable synchronization state is achieved if all the
oscillators of the system are in synchrony. This means that the
synchronized state defined by the $N$ constraints
$(x_1,y_1)=(x_2,y_2)=(x_3,y_3)=...=(x_N,y_N)$ (within some
precision or tolerance $h$) is in the synchronization manifold and
therefore the stability of the synchronization process is ensured.
The numerical synchronization criterion is defined as the sum of
the absolute
 deviations from each oscillator respect to all other, and dividing such sum for the number of terms:

\begin{eqnarray}
\label{eq7}
 \frac{1}{N(N-1)}\sum_{i\neq j}|x_i-x_j|<h, \qquad \qquad \forall (i,j).
\end{eqnarray}
It should be underlined that this numerical synchronization
criterion gives an insight about perfect synchronization in the
ring but does not tell anything about other types of  dynamical
behavior related to stable synchronization. In some regions where
stable synchronization is expected from MSF analysis, one can
instead observe  unstable synchronization in the full model (2).
The discrepancy  depends upon the choice of initial conditions and
the number of the self-sustained oscillators which are taken into
account during the process.
The neglected nonlinear terms give rise to nonlinear contributions that apparently increase with increasing the coupling $K$, thus explaining the discrepancy between the MSF method and the numerical simulations.
From the criterion (\ref{eq7}), we
have obtained some results that depend on the number $N$.
In particular we have noticed that increasing the number of oscillators the domain of stability of all modes shrinks. For instance, increasing the number of oscillators from $N=6$ to $N=8$ in Fig. 3(ii)  the Lyapunov exponents associated to each mode become positive for a lower value of the coupling constant $K$: the first mode becomes unstable for $K>-0.014$ when $N=6$, for $K>-0.019$ for $N=7$, and for $K>-0.024$ for $N=8$. An analogous behavior is observed for the other modes.

Both full synchronized states and no synchronization can be found in
Fig. 3, where we also display the comparison between analytical
and numerical results. One also finds some domains obtained only
from numerical simulation that do not match with the MSF stability
condition. It is important to notice that this gap depends upon
the choice of the precision or tolerance $h$. In fact, given the
accuracy of the numerical integration, if $h$ is too small
$(h<<10^{-6})$ one would never observe synchronization, while for
too a large $h$ $(h>>10^{-6})$ synchronization would appear as a
spurious effect, so it is necessary to determine an appropriate
value of the parameter $h$ as a function of the accuracy of the
numerical scheme,
a value that we have estimated for our scheme to be $h\simeq 10^{-6}$. To
circumvent such difficulty (and arbitrariness) we will suggest in Sect. IV
another tool to detect synchronization, the Kuramoto order parameter.

Figs. 4 and 5 show space-time-amplitude diagrams that display the
behavior for two values of the coupling parameter $K$ in the
unstable or no-synchronization (US) and stable synchronization
(SS) domains. For the unstable domain, Fig. 4 shows the time
evolution for $(N,K)=(30,-0.2)$ and $(N,K)=(50,-0.05)$, where no
synchronization is found in the ring. Considering the domain of
stable synchronization, we display in Fig. 5 for $(N,K)=(20,0.6)$
and $(N,K)=(50,1)$ the process of entrainment of the oscillators
in the ring.

\section{The ring of nonidentical coupled oscillators}

The understanding of the collective dynamics of synchronization
processes occurring in living organisms is a paramount task of formidable difficulty. One
reason is that in a realistic description of such systems, even
within the same species, oscillators are made of non identical
elements. Although not explicitly considered in this work, real
systems are also always subject to noise in the form of
fluctuations associated with dissipation, as well as in the form
of random force due to the  external environment. We will, however, focus
our attention on quenched disorder, or non identical oscillators that can be modelled
with a spread of the characteristic parameters.
For example, a ring of nonidentical oscillators is
useful to investigate calcium dynamics in pancreatic acinar cells
\cite{Tsaneva} as well as other complex networks \cite{Strogatz}.
In the case where Eq.(\ref{eq2}) is used to model biological
systems of immobilized enzymes, the interpretation of the
parameters $\mu$, $\alpha$ and $\beta$  requires that they can
vary from site to site, according to the tendency degree of the
oscillator to possess a ferroelectric behavior and also to the
conductivity of the medium \cite{enjieu-chabi-yamapi-woafo}. This
leads to the possibility to have nonidentical biological
oscillators, or to a ring of nonidentical coupled systems
described by the following equations, where $\alpha$ and $\beta$
are constant, while $\mu$ is not:

\begin{eqnarray}
\label{eq8} && \ddot x_1-\mu_1 (1- x^2_1+\alpha x_1^4-\beta
x_1^6)\dot x_1+w^2_1 x_1=K(
x_{2}- 2x_1+x_{N}), \nonumber\\
&& \ddot x_{\nu}-\mu_\nu (1- x^2_{\nu}+\alpha x_{\nu}^4-\beta
x_{\nu}^6)\dot x_{\nu}+w^2_\nu x_{\nu}=K( x_{\nu+1}-
2x_{\nu}+x_{\nu-1}), \qquad
 \nu=2,...,N-1,\nonumber\\
&&\ddot x_{N}-\mu_N (1- x^2_{N}+\alpha x_{N}^4-\beta x_N^6)\dot
x_N+w^2_N x_{N}=K( x_{1}- 2x_{N}+x_{N-1}).
\end{eqnarray}
We recall that $\mu_\nu$ are positive coefficients; they measure
the dissipative strength and  are smaller than unity. $w_\nu$ is
the frequency of the associate linear oscillator ($\mu_\nu=0$) and
depends upon the site. The nonlinear terms play the role of
amplitude-dependent dissipation and provide a self-sustaining
mechanism for the perpetual oscillation. For the original
biological system \cite{enjieu-chabi-yamapi-woafo,kaiser},
$\mu_\nu$ can be expressed as function of the parameters of the
systems as follows

\begin{eqnarray}
\label{eq9}
 \mu_\nu&=&\frac{\kappa^2-\sigma^2}{w_{o\nu}}=\frac{18}{5}\alpha w_{o\nu}\kappa^2,
\end{eqnarray}
where $w_{o\nu}$ are the frequencies of the periodic
enzyme-substrate reaction and can be determined by the
recombination and attraction coefficients $\gamma_\nu, \xi_\nu$ as
$w_{o\nu}=\sqrt{\gamma_\nu \xi_\nu}$
\cite{enjieu-chabi-yamapi-woafo}. $\xi_\nu$ is the decay rate of
each excited enzymes to the ground (or weakly polar) state and
$\gamma_\nu$ the range attraction of the substrate particles due
to the autocatalytic reactions. From Fr\"olich ideas, we may
suppose that in large regions of the system of proteins,
substrates, ions and structured water are activated by the
chemical energy available from substrate enzyme reactions
\cite{frolich}. Thus, chemical oscillations in the number of
substrate and activated enzyme molecules with a very low frequency
$w_{o\nu}$ might be carried out around the equilibrium state
\cite{kaiser}. $\kappa$ (obtained through  a nonlinear dielectric
contribution) is the coefficient proportional to the macroscopic
polarization and the time dependent number of the excited enzyme
molecules \cite{kaiser}. It can be chosen in the interval
$1<\kappa<5$. $\sigma$ is viewed as a coefficient of relaxation
term of electric resistances against the system's tendency to
become ferroelectric. We assume that the frequency  $w_{o\nu}$ is
chosen as: $w_{o\nu}=1+ \Delta w_o (\zeta_\nu-\frac{1}{2})$
($\zeta_\nu$ is a uniform random variable less than the unity and
$\Delta w_o$ a disorder parameter). We also assume that the
natural frequencies $w_{\nu}$ are uniformly distributed around
$1$: $w_{\nu}=1+ \Delta w (\zeta_\nu-\frac{1}{2})$. For
simplicity, to avoid a new parameter, we take the spread of the
recombination frequencies $\Delta w_o$ and of the natural
frequencies $\Delta w$ to be the same: $\Delta w_o = \Delta w$.
 According to the
analytical expression the amplitudes $A$ and frequencies
$\Omega_\nu$ of the limit-cycles established in
Ref.\cite{enjieu-chabi-yamapi-woafo},  the periodic solution for
each oscillator is approximately described  by
\begin{eqnarray}
\label{eq10} x_\nu(t)=A\cos\Omega_\nu t
 \end{eqnarray}
where the amplitude $A$ and the frequencies  $\Omega_\nu $ for the
MLC-vdPo (\ref{eq1}) are derived through the following equations:
\begin{eqnarray}
\label{eq11}
&&\frac{5\beta}{64}A^6-\frac{\alpha}{8}A^4+\frac{1}{4}A^2-1=0, \hspace{9cm} 11(a)\nonumber\\
&&\Omega_\nu=w_\nu+\frac{\mu^2}{w_\nu}\left \{
\frac{1580\beta}{393216}A^{12}-\frac{738\alpha\beta}{99024}A^{10}+(\frac{72\alpha^2+309\beta}{768})A^8-(\frac{64\alpha-219\beta}{6144})A^6
\right. \nonumber \\&& \left.+
(\frac{16\alpha+3}{384})A^4-\frac{3}{64}A^2 \right \}
+O(\mu^3).\hspace{8cm} 11(b)\nonumber
\end{eqnarray}
We note that only the frequencies $\Omega_\nu$ are modified by the
disorder parameter $\Delta w_o$, while the amplitude $A$ of the
orbit does not change. Thus, each self-sustained system in the
ring exhibits a different frequency, the amplitude, see Eq.
(\ref{eq11}(a)), only depends upon the $\alpha$ and $\beta$
parameters, which are fixed here. The self-sustained oscillators
with nonidentical $\mu_\nu$ consist therefore of oscillators with
nonidentical natural frequencies $\Omega_{\nu},$ that are a
function of the physical parameters $ \mu_\nu$, $\alpha$, $\beta$,
while the amplitude  $A$ of the limit-cycles is unchanged
\cite{enjieu-chabi-yamapi-woafo}. Therefore in the uncoupled limit
approximated by the harmonic oscillations, see Eq.(10), each
oscillator is completely described by a phase $\delta_{\nu}
=\cos^{-1} \left( x_{\nu}(t)/A\right)$ and  amounts to a rotator.
It is therefore tempting to monitor the rotators through the
phases $\delta_{\nu}$ analogously to the Kuramoto model
\cite{kuramoto,acebron,arenas08}, a paradigm for coupled rotators
and already employed to investigate globally coupled van der Pol
oscillators \cite{peles03}. Synchronization of the nonidentical
oscillators (8) can be revealed by means of the Kuramoto order
parameter $R$ defined as \cite{kuramoto,acebron,arenas08}:

\begin{eqnarray}
\setcounter{equation}{12}
\label{kuramoto}
R=\frac{1}{N}\sum_{j=1}^N e^{i\delta_j}.
\label{kuramotoorderparam}
\end{eqnarray}
For a completely disordered steady-state one gets (for very large
$N$) $R \simeq 0$, while $R \simeq 1$ corresponds to phase
synchronization. The advantage to employ such parameter $R$ is
that it is possible to measure the degree of synchronization, or,
roughly speaking, the fraction of synchronized oscillators in the
intermediate cases of partial synchronization, for which the
parameter $R$ will reach a value $0<R<1$. So the Kuramoto order
parameter $R$ is capable to describe also states where condition
(\ref{eq7}) is not satisfied, but nevertheless the oscillators
exhibit some tendency to behave coherently. The interest to
determine the tendency of nonidentical systems to behave
synchronously is almost ubiquitous, from laser arrays
\cite{brusselbach} to people walking on a bridge \cite{Mcrobie},
or Josephson arrays \cite{daniels}. For instance in the case of
Josephson oscillators it has been speculated that the analysis of
the Kuramoto order parameter (\ref{kuramoto}) is useful to predict
the fraction of coherently working electronic elements, and
therefore the efficiency of the devices
\cite{wiesenfeld-colet-strogatz,filatrella-eurphysjb,dhamala-wiesenfeld}.
Here we will use the Kuramoto order parameter in the same spirit:
as a diagnostic tool to measure the capability of the oscillators
to synchronize in spite of the differences in the single
oscillators frequencies. In order to look at the qualitative
picture of the dynamical states of oscillators under the influence
of diffusive coupling, we have plotted in Fig. 6 the Kuramoto
order parameter as a function of the coupling parameter $K$ for
several different finite number $N$ of oscillators, as analyzed
for instance in Ref. \cite{pazo}. From Fig. 6, it can be noticed
that full synchronization occurs mainly for $K>0$, as expected
from MSF analysis, although the Kuramoto order parameter allows an
analysis of the partial synchronization
( corresponding to a state - analogous to clustering - where the maximum Lyapunov exponent is positive, but not all modes are unstable  \cite{xie})
that occurs in the region where full synchronization is unstable, see Sect. III.
Therefore, this finding suggests that in the case of non-identical
oscillators, excitatory coupling is more suitable to achieve
coherent oscillations when compared to the inhibitory coupling
($K<0$). It can be notice that synchronization in presence of
disorder is not achieved for negative $K$ values  as we show on
Fig.7 for an enlargement around low $K$ values. From the Kuramoto
order parameter (\ref{kuramoto}), the following results have been
obtained for $N=6$, $N=10$, $N=20$ and $N=30$. When the ring is
composed of $N=6$ nonidentical coupled oscillators, the phase
synchronization state is achieved for $K\geq 1.6$. As the number
of nonidentical oscillators in the ring increases, one still
observes two main dynamical states but for different ranges of
$K$, and in general the degree of synchronization decreases when
the number of oscillators is increased. In the case of a ring is
composed of $N=10$ and $20$ nonidentical oscillators, full
synchronization is obtained approximately in the range $K>3.6$ and
$K>10.5$, respectively. For $N=30$ oscillators in the ring, the
oscillators are fully synchronized only for $K >26.0$.

It is also interesting to compare the result of the Kuramoto order
parameter analysis in Fig. 7 with the MSF analysis of Fig. 3. In
this enlargement it is clear that disorder ($\Delta w_o=0.05$)
induces an enlargement of the desynchronization region around
$K=0$. For instance for $N=10$ while the uniform oscillators are
not synchronized only in the narrow region $-0.028<K<0.005$, after
which there is a range of perfect synchronization (see also Table
1), from the simulation of disordered oscillators in Fig. 7 it is
evident that there is a much more extended region of partial
synchronization. Nevertheless it is clear that the qualitative
behavior is the same in Fig. 3 and Fig. 6: the region unfavorable
to synchronization around low $K$ values clearly increases
increasing the number $N$ of oscillators in both the ordered and
disordered cases.

The behaviors of the space-time-amplitude in the partially
synchronized state and in the fully synchronous motion are similar
to the patterns shown in Figs. 4 and 5. The ring of nonidentical
oscillators capability to give rise to synchronization is
represented in Table 2.

 {\scriptsize
\begin{eqnarray*}
\begin{tabular}{|c|c|c|}
\hline $N$&Domains of K for synchronization (i)&Domains of K for synchronization (ii)\\
\hline  2& [0.4;+$\infty$ [&  [0.2;+$\infty$ [\\
\hline  3& [0.6;+$\infty$ [&  [0.3;+$\infty$ [\\
\hline  4& [0.7;+$\infty$ [ &  [0.6;+$\infty$ [\\
\hline  5& [1.0;+$\infty$ [&  [0.7;+$\infty$ [\\
\hline 6 & [1.3;+$\infty$ [&  [0.8;+$\infty$ [\\
\hline 8 & [2.4;+$\infty$ [&  [1.3;+$\infty$ [\\
\hline 10 &[3.6;+$\infty$ [&  [1.6;+$\infty$ [\\
\hline  15& [7.2;+$\infty$ [& [4.1;+$\infty$ [\\
\hline  20& [12.8;+$\infty$ [&  [7.6;+$\infty$ [\\
\hline  25& [19.4;+$\infty$ [&  [ 10.6;+$\infty$ [\\
\hline  30& [26.0;+$\infty$ [& [14.8;+$\infty$ [\\
\hline
\end{tabular}
\end{eqnarray*}
\begin{center}
{\it Table 2: Domains of full ($R>0.998$) synchronization in the
ring of nearest neighbors coupled nonidentical oscillators.
Parameters of the system are (i) $\mu=0.1$, $\alpha=0.1$,
$\beta=0.1$; (ii) $\mu=0.1$, $\alpha=0.114$, $\beta=0.005$. The
spread of the frequencies is $\Delta w_o=0.05$.}
 \end{center}}

 Figure 8 shows the stability diagram in the $(N,K)$
 plane and reveals the boundary between phase synchronization (defined as $R>0.998$) and
 disordered steady-state. These curves also show that the dynamical
 states depend on the ring size.
Analyzing the effects of the disorder parameter $\Delta w_o$ (that
controls the spread of the frequencies of the uncoupled
oscillators) on various dynamical states in the ring, one finds
that such disorder parameter can also induce de-synchronization,
{\it i.e.} even preparing the oscillators with identical initial
conditions, the spread induces a disordered state.

\section{An open-ended chain of nearest neighbors coupled oscillators }
\subsection{A chain of open-ended  nearest neighbors coupled identical oscillators}

{\scriptsize
\begin{eqnarray*}
\begin{tabular}{|c|c|c|}
\hline $N$&Domains of $K$ for stable synchronization (i)&Domains of $K$ for stable synchronization (ii)\\
\hline  2&  ]-0.5;-0.01[$\cup$]0.014;+$\infty$ [&  ]-0.5;-0.01[$\cup$]0.009;+$\infty$ [\\
\hline  3& ]-0.34;-0.02[$\cup$]0.02;+$\infty$ [&  ]-0.334;-0.019[$\cup$]0.019;+$\infty$ [\\
\hline  4& ]-0.3;-0.03[$\cup$]0.04;+$\infty$ [ &  ]-0.3;-0.03[$\cup$]0.03;+$\infty$ [\\
\hline  5& ]-0.28;-0.05[$\cup$]0.07;+$\infty$ [&  ]-0.28;-0.05[$\cup$]0.05;+$\infty$ [\\
\hline 6 & ]-0.27;-0.07[$\cup$]0.1;+$\infty$ [&  ]-0.27;-0.07[$\cup$]0.07;+$\infty$ [\\
\hline 8 & ]-0.26;-0.13[$\cup$]0.18;+$\infty$ [&  ]-0.26;-0.13[$\cup$]0.12;+$\infty$ [\\
\hline 10 &  ]-0.26;-0.21[$\cup$]0.29;+$\infty$ [&  ]-0.26;-0.21[$\cup$]0.19;+$\infty$ [\\
\hline  20& ]1.16;+$\infty$ [& ]0.77;+$\infty$ [\\
\hline  30& ]2.61;+$\infty$ [&  ]1.74;+$\infty$ [\\
\hline  40& ]4.66;+$\infty$ [&  ] 3.1;+$\infty$ [\\
\hline  50&  ]7.28;+$\infty$ [& ]4.85;+$\infty$ [\\
\hline
\end{tabular}
\end{eqnarray*}
\begin{center}
{\it Table 3: Domains of stable synchronization in the open-ended
chain of nearest neighbors coupled identical oscillators obtained
with the MSF approach. Parameters of the system are (i):
$\mu=0.1$, $\alpha=0.1$, $\beta=0.1$, (ii) $\mu=0.1$,
$\alpha=0.114$,$\beta=0.005$.}
 \end{center}}

Not closed chains of coupled nonlinear systems, corresponding to
open-ended boundary condition,
 are involved in many natural
processes such as the swimming motion of organisms
\cite{Williams-1992} and waves synchronization that occur during
sensory processing in the cortex \cite{Wilson-1992,Gray-1989}. In
the case of small intestinal muscle, it has proved useful to
simulate
 and attempt an analysis of long chains of oscillators \cite{Brown}. There is
 in fact physiological evidence that a complete organ such
  as small intestine, comprises a very large
number of smooth muscle cells organized to form self-oscillatory
segments. Also, the electrical waveforms found in the canine
gastrointestinal tract are very non-sinusoidal in nature, whereas
research work realized in the Department of Surgery, University of
Sheffield has indicated that the human duodenal signals are nearly
sinusoidal \cite{Brown-1971,Duthier-1971,Duthier-1972}.
Consequently, the nonlinear parameter $\mu$ can be assumed small
to model the duodenal oscillator cells. In fact in the autonomous
regime, the final limit cycle state of the van der Pol oscillator
is nearest  a sinusoidal behavior for small values of $\mu$. In
contrast, for large $\mu$ it develops relaxation oscillations
\cite{Van-1922,Van-1934}. In the open-ended chain coupling
configuration, one assumes that the $1^{th}$ and $N^{th}$
oscillators are not interconnected.  In this case, the
differential equations of motion are defined as

\begin{eqnarray}
\label{eq13} && \ddot x_{1}-\mu (1- x^2_{1}+\alpha x_{1}^4-\beta
x_{1}^6)\dot
x_{1}+x_{1}=K(x_2-x_1),\nonumber\\
 && \ddot x_{i}-\mu (1- x^2_{i}+\alpha
x_{i}^4-\beta x_{i}^6)\dot
x_{i}+x_{i}=K(x_{i+1}-2x_i+x_{i-1}),\qquad i=2,3,...,N-1\nonumber\\
 && \ddot x_{N}-\mu (1-
x^2_{N}+\alpha x_{N}^4-\beta x_{N}^6)\dot
x_{N}+x_{N}=K(x_{N-1}-x_N).
\end{eqnarray}
Our purpose is to recover the essential features of the stability
investigation (employing the same methods of the section III and
IV) to identify some general properties of an open-ended chain of
nearest neighbors coupled self-sustained systems. The stability of
the synchronization process is again found following the MSF
approach. In the present case, the coupling matrix is
\begin{displaymath}
\mathbf{ G_{OE} }=\left (
\begin{array}{ccccc}
-1&1& 0& \ldots &0\\
1 &-2&1& \ldots &0\\
0&1&-2&\ldots&0\\
\vdots &\vdots&\vdots&\ddots&\vdots\\
0&0&\ldots &1&-1
\end{array}
\right ).
\end{displaymath}

{\scriptsize
\begin{eqnarray*}
\begin{tabular}{|c|c|c|}
\hline $N$&Domains of K for stable synchronization in the ring&Domains of K for stable synchronization in the open-ended chain\\
\hline  10&  [-0.24;-0.03]$\cup$[0.06;+$\infty$ [&  ]-0.25;-0.22[$\cup$]0.3;+$\infty$ [\\
\hline  20& [-0.24;-0.12]$\cup$[0.22;+$\infty$ [&  ]1.17;+$\infty$ [\\
\hline  30& [0.49;+$\infty$ [&  [2.62;+$\infty$ [\\
\hline  40& [0.87;+$\infty$ [&  [ 4.66;+$\infty$ [\\
\hline  50&  [1.35;+$\infty$ [& [7.28;+$\infty$ [\\
\hline
\end{tabular}
\end{eqnarray*}
\begin{center}
{\it Table 4: Comparison between stability domains in the ring and
the open-ended chain of identical oscillators obtained with the
MSF  approach, case (i): $\mu=0.1$, $\alpha=0.1$, $\beta=0.1$.}
 \end{center}}

The coupling matrix $\mathbf{ G_{OE} }$ also obeys  the zero sum
condition, and the synchronization manifold $\mathcal{M}$ is an
invariant set. Therefore, stability of the synchronous state
reduces to find the behaviors of the system dynamical properties
along directions in phase that are transverse to the
synchronization manifold. The stability of the resulting dynamical
states can be determined considering the set of variational
equations (6). In Ref. \cite{pecora-1998}, it was  showed that the
matrix $\mathbf{ G_{OE} }$ can be diagonalized in a manner similar
to the shift-invariant case using a discrete Fourier transform
with N replaced by $2N$, and the eigenvalues are
$\gamma_k=-4\sin^2\left ( \pi k/2N \right ),$ for $k=0,1, . . .
,N-1$. This is similar to  the case of diffusively coupled
self-sustained systems, except that there are no degenerate modes
and the highest wavelength corresponds to $k=N-1$. Nonetheless,
because of the dependence of the eigenvalues on $N$, we will see
the same dynamical states as before. The stability boundaries of
the synchronization  derived through the MSF  are shown in the
Table 3:

 The dependence of the size of a
chain also apparent in Fig. 9, where the MSF is plotted in the
plan $(K,N)$ for the open-ended chain. It appears that,
analogously to the results reported in Fig. 2, as the size of a
chain increases, the domain of stable synchronization reduces more
quickly than in the case of the closed end model, see Table 4.

In Fig. 10 we plot the stability diagram in the $(N,K)$ plane for
the open-ended chain of coupled self-sustained oscillators,
obtained through the MSF and a direct numerical simulations of the
differential equations (\ref{eq13}). Comparison with the case of
the ring, (see Fig. 3), reveals that the agreement between the
results obtained from the MSF and those of a direct numerical
simulation is very good. It also indicates that  the region of
stable synchronization is smaller then in the case of the ring, as
reported in Tables 4 and 5.

{\scriptsize
\begin{eqnarray*}
\begin{tabular}{|c|c|c|}
\hline $N$&Domains of K for stable synchronization in the ring&Domains of K for stable synchronization in the open-ended chain\\
\hline  10&  [-0.24;-0.04]$\cup$[0.04;+$\infty$ [&  ]-0.25;-0.21[$\cup$]0.2;+$\infty$ [\\
\hline  20& [-0.24;-0.12]$\cup$[0.13;+$\infty$ [&  ]0.78;+$\infty$ [\\
\hline  30& [0.29;+$\infty$ [&  [1.75;+$\infty$ [\\
\hline  40& [0.51;+$\infty$ [&  [ 3.11;+$\infty$ [\\
\hline  50&  [0.79;+$\infty$ [& [4.86;+$\infty$ [\\
\hline
\end{tabular}
\end{eqnarray*}
\begin{center}
{\it Table 5: Comparison between stability domains in the ring and
the open-ended chain of coupled identical oscillators obtained
with the MSF approach, case (ii): $\mu=0.1$,
$\alpha=0.114$,$\beta=0.005$..}
 \end{center}}

Finally, let us add that the space profiles and the time dependent
amplitudes in both cases of stable and unstable (or partially
synchronized) states are very similar to those shown in Figs. 4
and 5, and therefore are not shown.

\subsection{A chain of open-ended  nearest neighbors coupled non-identical oscillators}
 \noindent

In the case of an open-ended chain of nearest neighbors coupled
non-identical oscillators, the chain's dynamics is given by the
following set of equations:
\begin{eqnarray}
\label{eq14} && \ddot x_{1}-\mu_1 (1- x^2_{1}+\alpha x_{1}^4-\beta
x_{1}^6)\dot
x_{1}+w^2_1x_{1}=K(x_2-x_1),\nonumber\\
 && \ddot x_{i}-\mu_i (1- x^2_{i}+\alpha
x_{i}^4-\beta x_{i}^6)\dot
x_{i}+w^2_i x_{i}=K(x_{i+1}-2x_i+x_{i-1}),\qquad i=2,3,...,N-1,\nonumber\\
 && \ddot x_{N}-\mu_N (1-
x^2_{N}+\alpha x_{N}^4-\beta x_{N}^6)\dot
x_{N}+w^2_Nx_{N}=K(x_{N-1}-x_N).
\end{eqnarray}
Our choice of considering non-identical coefficients is justified
by the parameter spread encountered in real systems, as discussed
in section IV for the ring of nearest neighbors coupled non
identical self-sustained oscillators. The set of equations (\ref{eq14}) is numerically
integrated  to compute the Kuramoto order parameter $R$,
Eq.~(\ref{kuramoto}), versus the coupling coefficient $K$, to
quantify the fraction of synchronized oscillators dynamics in the
chain of open-ended nearest neighbors coupled non-identical
oscillators.

Fig. 11 shows the variation of the Kuramoto order
parameter $R$ versus the coupling coefficient $K$ for $N=6$,
$N=10$, $N=20$ and $N=30$. Phase synchronization is found in the
open-ended chain and the stability boundary depends on $N$.
However, as the number of oscillators $N$ increases, the open
ended chain is very sensitive to the initial conditions and this
dependence more pronounced when the coupling strength $K$ becomes
larger.  Fig.12 shows the stability diagram for an open-ended
chain of coupled nonidentical oscillators in the $(K,N)$ plane,
and it is qualitatively very similar to Fig. 8. The chain of
coupled nonidentical oscillators capability to give rise to
synchronization (defined again as $R>0.998$) is appears in Fig. 12
represented in Table 6 for both the sets of parameters.

\section{Conclusions}
\noindent

We have studied some criteria under which the synchronization
manifold is stable for a ring and an open-ended chain of nearest
neighbors coupled van der Pol-like oscillators. By using the
Master Stability Function, the stability boundaries of the main
synchronized states have been investigated and the obtained
results have been complemented by numerical simulations. Thereby,
the following findings have been captured: the threshold of the
coupling strength from which the system displays a stable
synchronized behavior is different for the cases of two sets of
parameters, as depicted in Fig. 2. Moreover, as far as the
diffusive coupling is concerned, we have found that with just the
first and last mode's behavior one can characterize the stability
of the system (as far as full synchronization is considered, {\it
i.e.} neglecting clustered states). The
stability boundaries of the
synchronization have been derived as a function of the number of
oscillators and of the coupling strength. We have found that
increasing the number of oscillators, the synchronized states are
unfavored, while, quite obviously, a larger coupling constant
increases their stability. The non obvious result is the presence
of a stable synchronization manifold even for {\it negative }
values of the coupling $K$, {\it i.e.} for repulsive {\it interaction}.
We have also found
that, when the size of the ring and the coupling coefficient
increases, the domains of instability become very large and the
gap between analytical and numerical analysis becomes more
relevant.

In the case of non identical oscillators, we have performed
numerical simulations to compute the fraction of entrained
oscillators by means of the Kuramoto order parameter. Varying the
coupling strength, we have retrieved the region of unstable and
stable synchronization states of the ring and an open ended chain
of non-identical nearest neighbors coupled self-sustained
oscillators. In particular, we have found that synchronization is
robust enough to survive also in presence of some amount of disorder. However, the
process depends upon the number of oscillators and the amount of disorder.

Research might be extended in several directions. For example we
think that an extension of the analytic treatment to find
stability of synchronous states in the ring of identical and non
identical coupled oscillators in the presence of noise is an
interesting task which can be tackled also using the Kuramoto
order parameter. Also, we have restricted our research to a region
of the
 parameters where only monorhythmical states are present. An interesting
 question is therefore the influence of two stable orbits on the synchronization
  properties in presence of disorder. Moreover, the spontaneous transition frequencies in the
ring with external random excitation, such as noise, can be also
experimentally observed if one can resolve in time enzymatic
reactions or electronic circuits.

\section*{Acknowledgements}
R.YAMAPI undertook this work with the support of the ICTP
(International Centre for Theoretical Physics) Programme for
Training and Research in Italian Laboratories, Trieste, Italy. He
also acknowledges the support of the Laboratorio Regionale
CNR/INFM at the  Dipartimento di Fisica, Universit\`a di Salerno
(Italy).

\newpage

\newpage



\newpage

\begin{figure}[htb]
\centering
\begin{center}
\begin{picture}(150,140)
\put(-40,-70.0)
{\includegraphics[width=18cm,height=20cm]{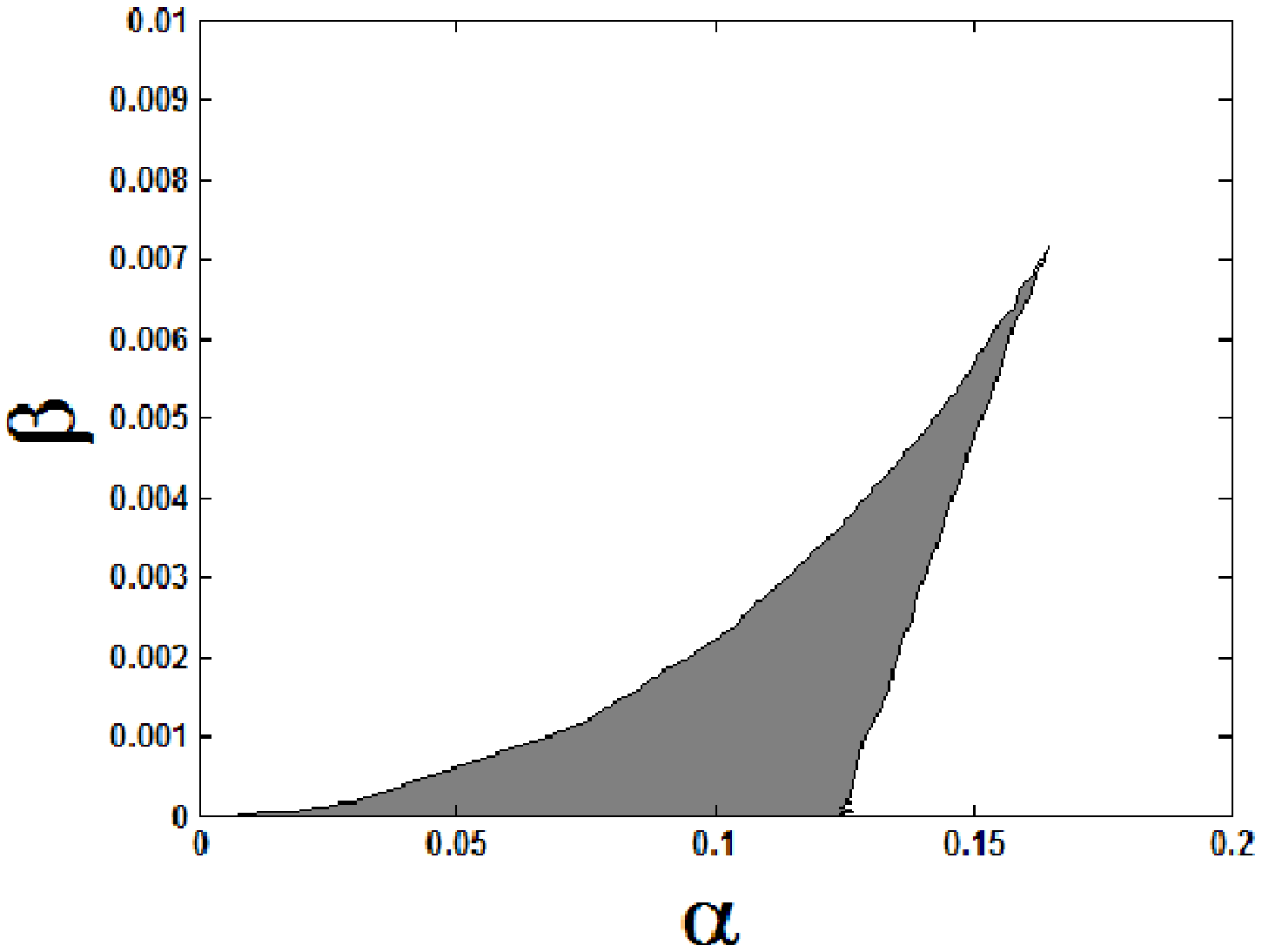}}
\end{picture}
\caption[] {    \it Parameters domain for the existence
of a single limit cycle (white area) and three limit cycles (grey
area) for $\mu=0.1$.} 
\end{center}
\end{figure}

\begin{figure}[htb]
\centering
\begin{center}
\begin{picture}(120,150)
\put(0,75)
{\includegraphics[width=12cm,height=7cm]{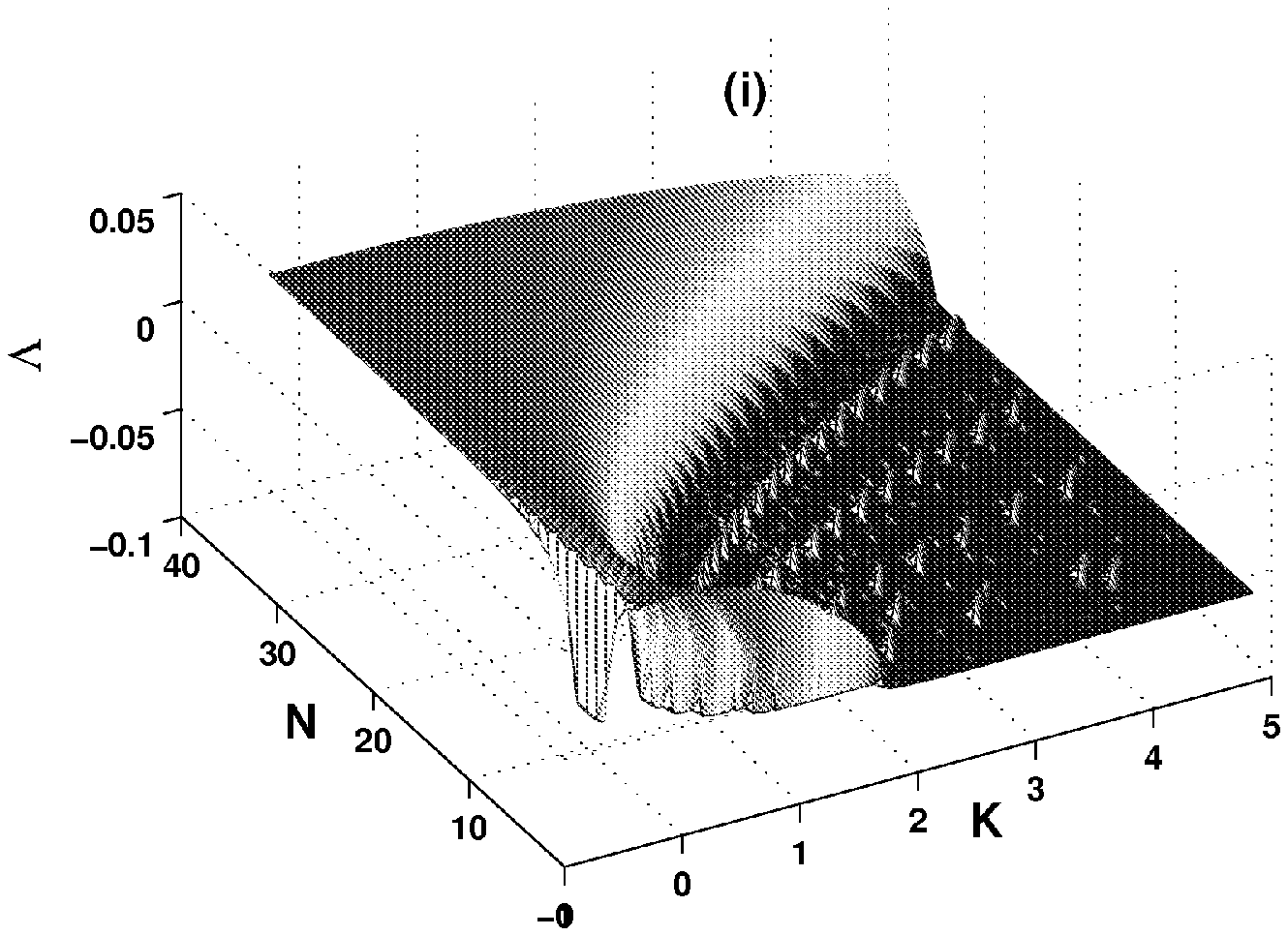}}
\put(0,0)
{\includegraphics[width=12cm,height=7cm]{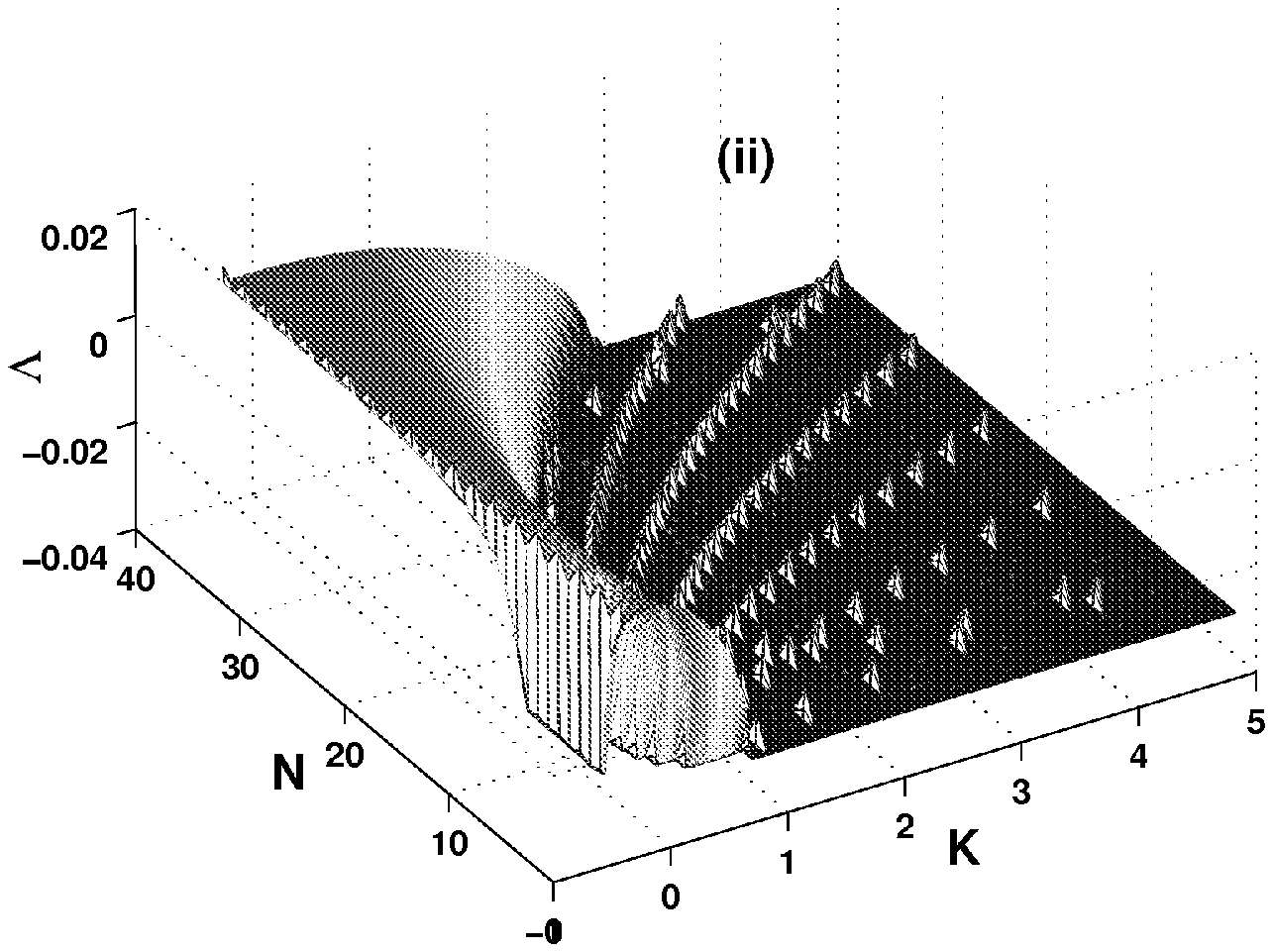}}
\end{picture}
\caption[] {    \it  Variation of the Master Stability
Function (MSF) $\Lambda$ versus $K$ and $N$ for a ring of
diffusive coupling oscillators, (i) $\mu=\beta=\alpha=0.1$ and (ii) $\mu=0.1; \beta=0.005, \alpha=0.114.$} 
\end{center}
\end{figure}

\begin{figure}[htb]
\centering
\begin{center}
\begin{picture}(250,150)
\put(0.0,75.0)
{\includegraphics[width=12.3cm,height=7cm]{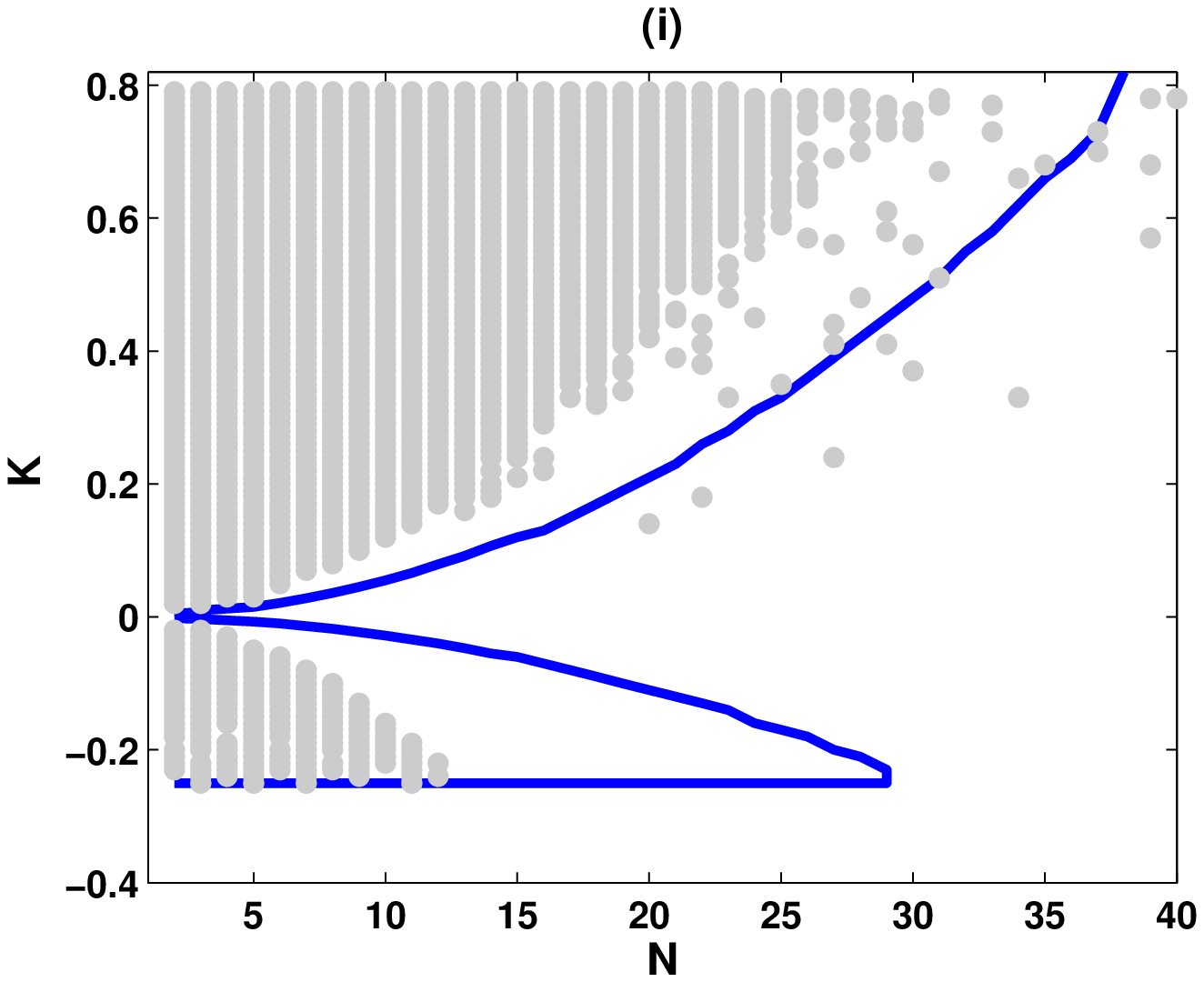}}
\put(0.0,0.0)
{\includegraphics[width=12.3cm,height=7cm]{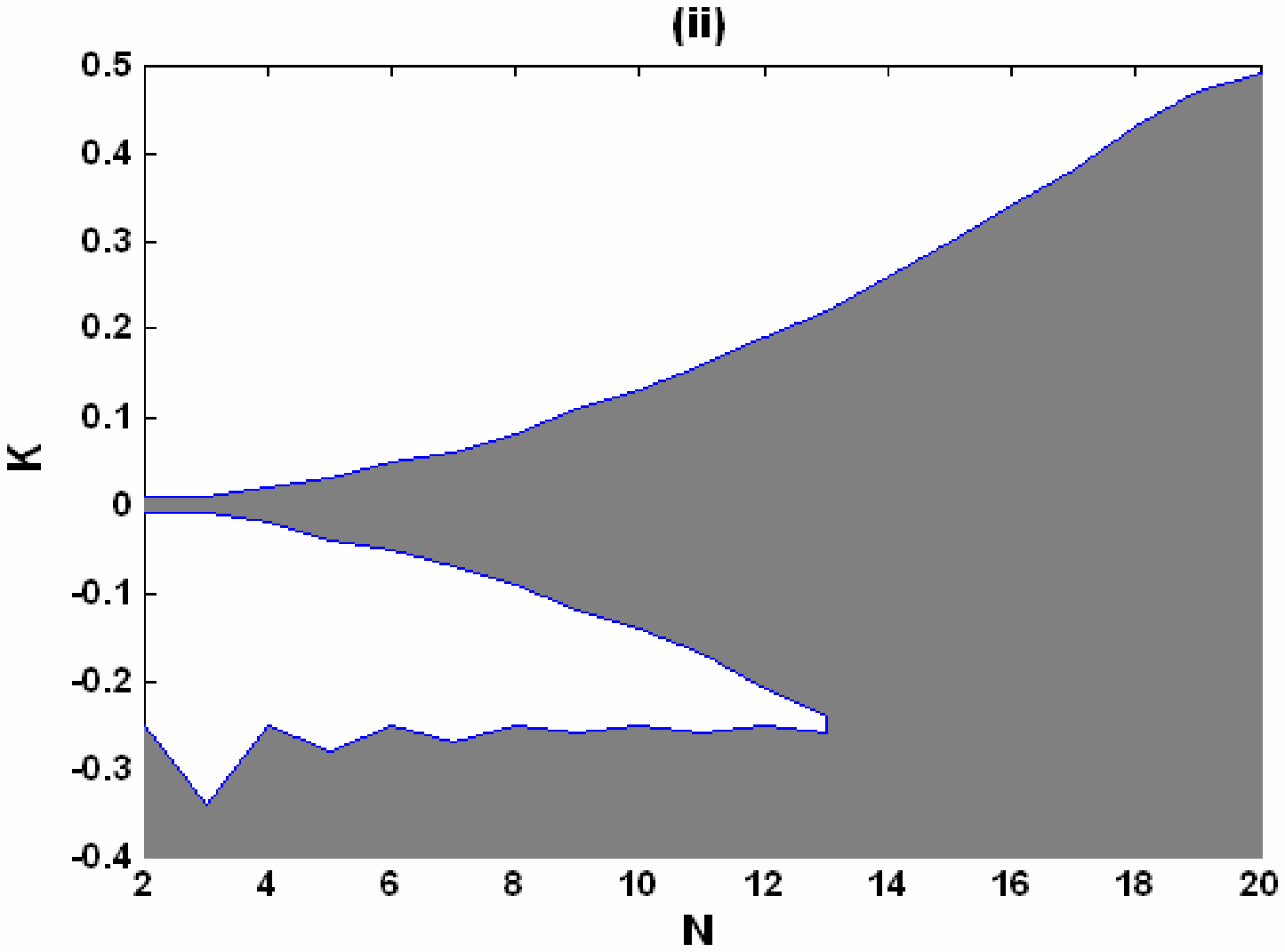}}
\end{picture}
\caption[] {    \it
Stability diagram in the
$(N,K)$ plane of a ring of diffusivly coupled oscillators.
The shaded area denotes the region of stable synchronization obtained  numerically with $h=10^{-6}$ (see Eq.(\ref{eq7}) ),
while the solid line is the stability boundary obtained through MSF.
Parameters of the system are $\mu=0.1$; (i) $\beta=\alpha=0.1$,
(ii) $\beta=0.005,\alpha=0.114$.
}
\end{center}
\end{figure}

\begin{figure}[htb]
\centering
\begin{center}
\begin{picture}(150,150)
\put(0.0,0.0)
{\includegraphics[width=12cm,height=7cm]{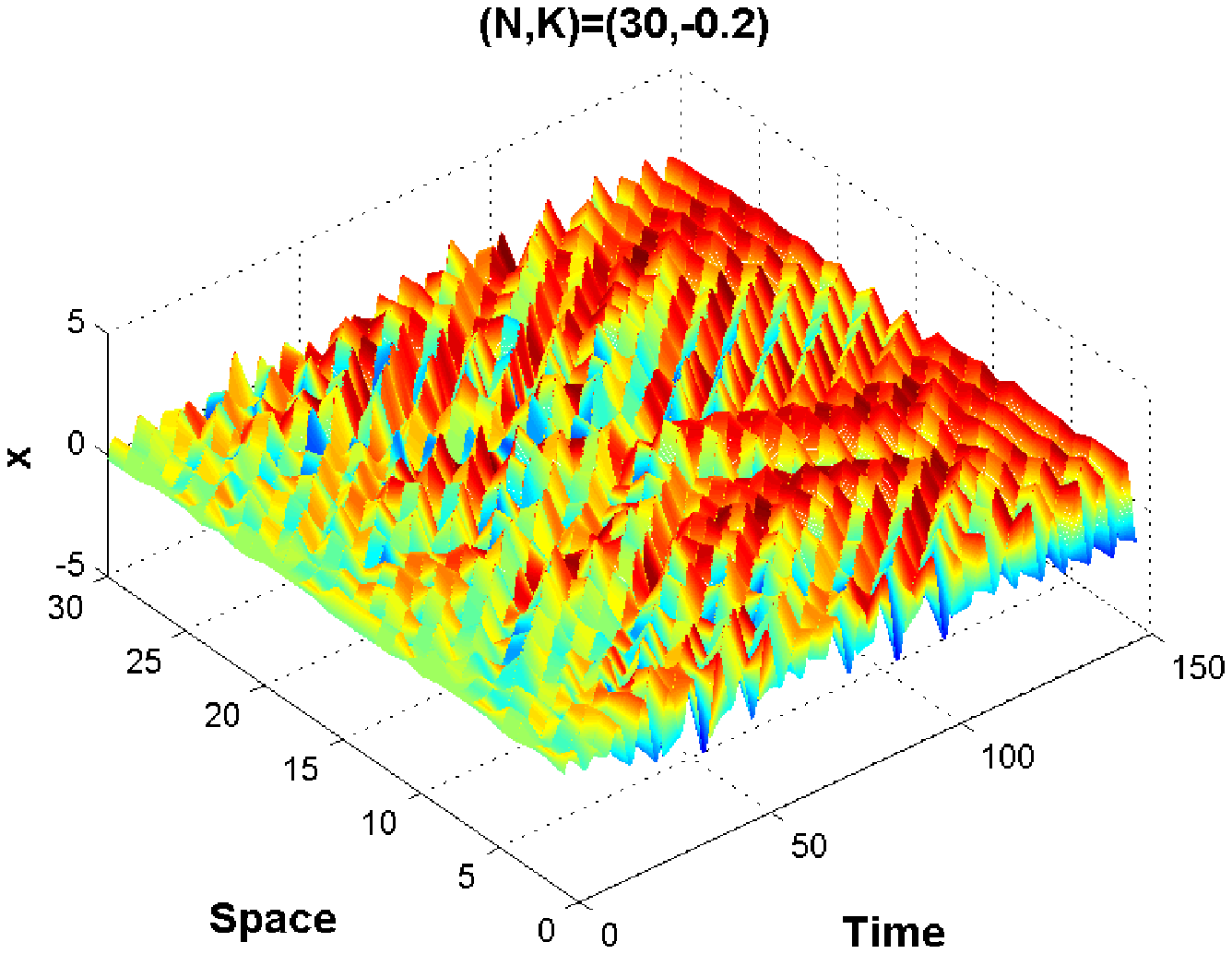}}
\put(0.0,70.0)
{\includegraphics[width=12cm,height=7cm]{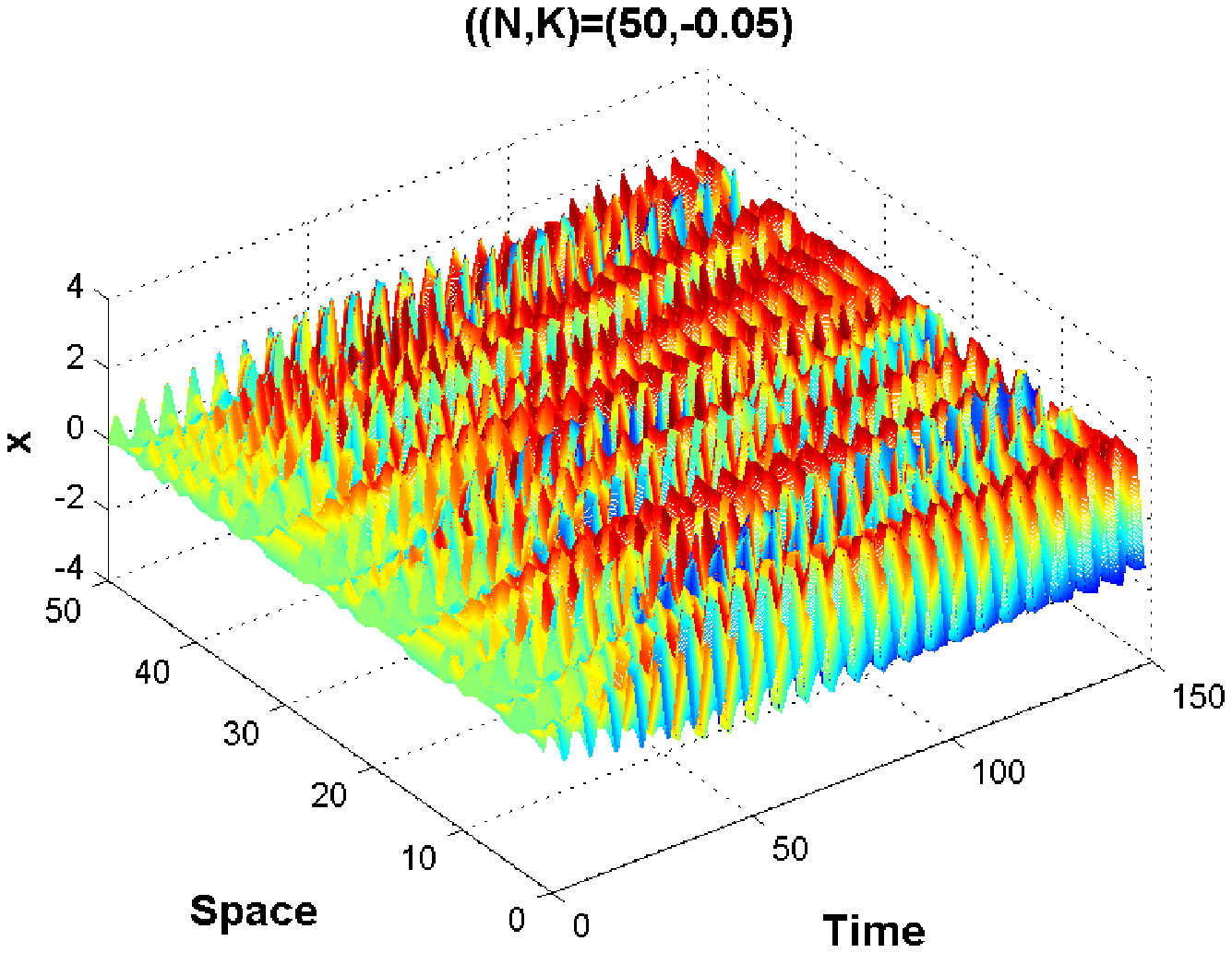}}
\end{picture}
\caption[] {    \it  Space-time-amplitude plot showing
unstable synchronization in the ring of diffusively coupled
oscillators, with $\mu=\alpha=\beta=0.1$. }
\end{center}
\end{figure}

\begin{figure}[htb]
\centering
\begin{center}
\begin{picture}(150,150)
\put(0.0,0.0)
{\includegraphics[width=12cm,height=7cm]{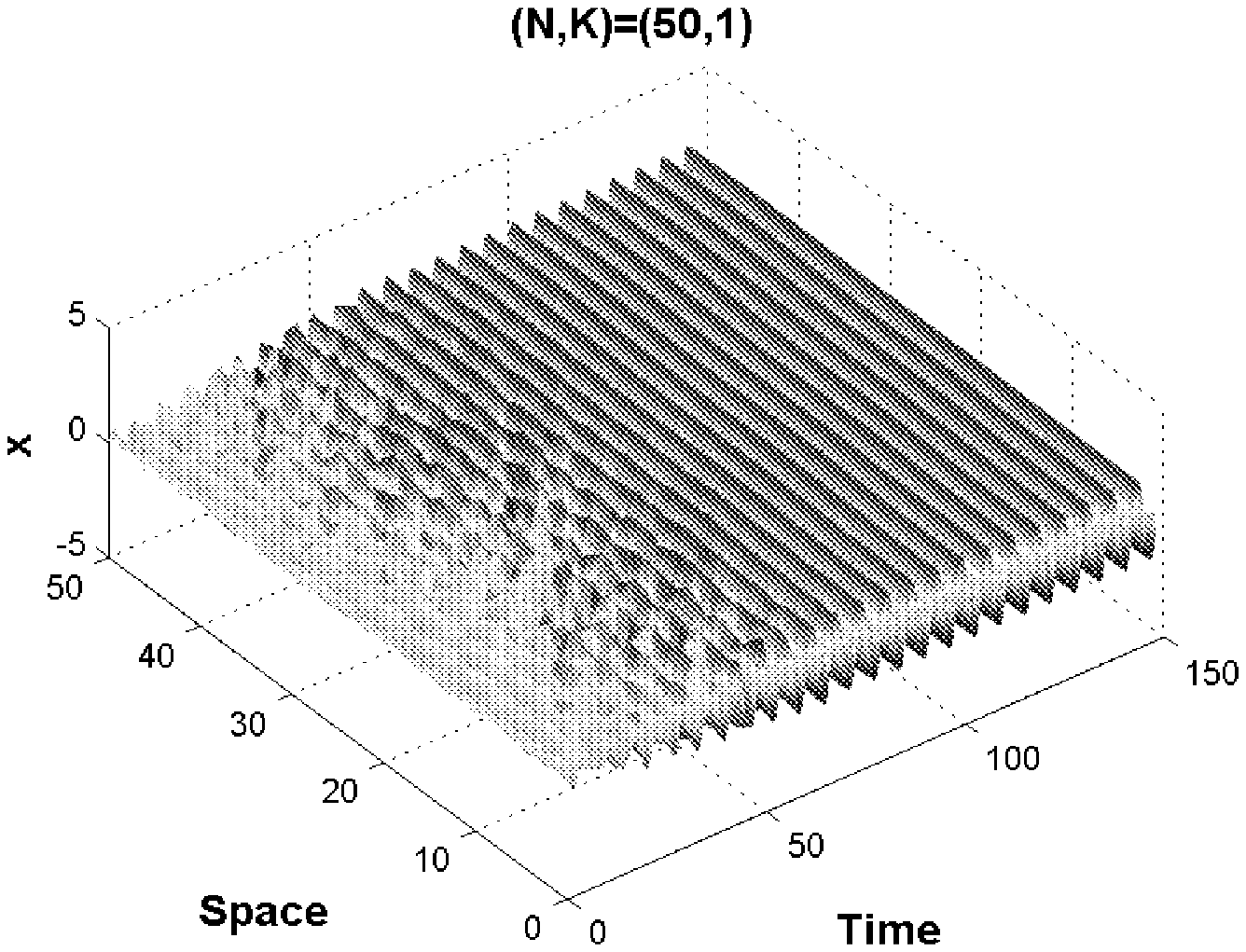}}
\put(0.0,70.0)
{\includegraphics[width=12cm,height=7cm]{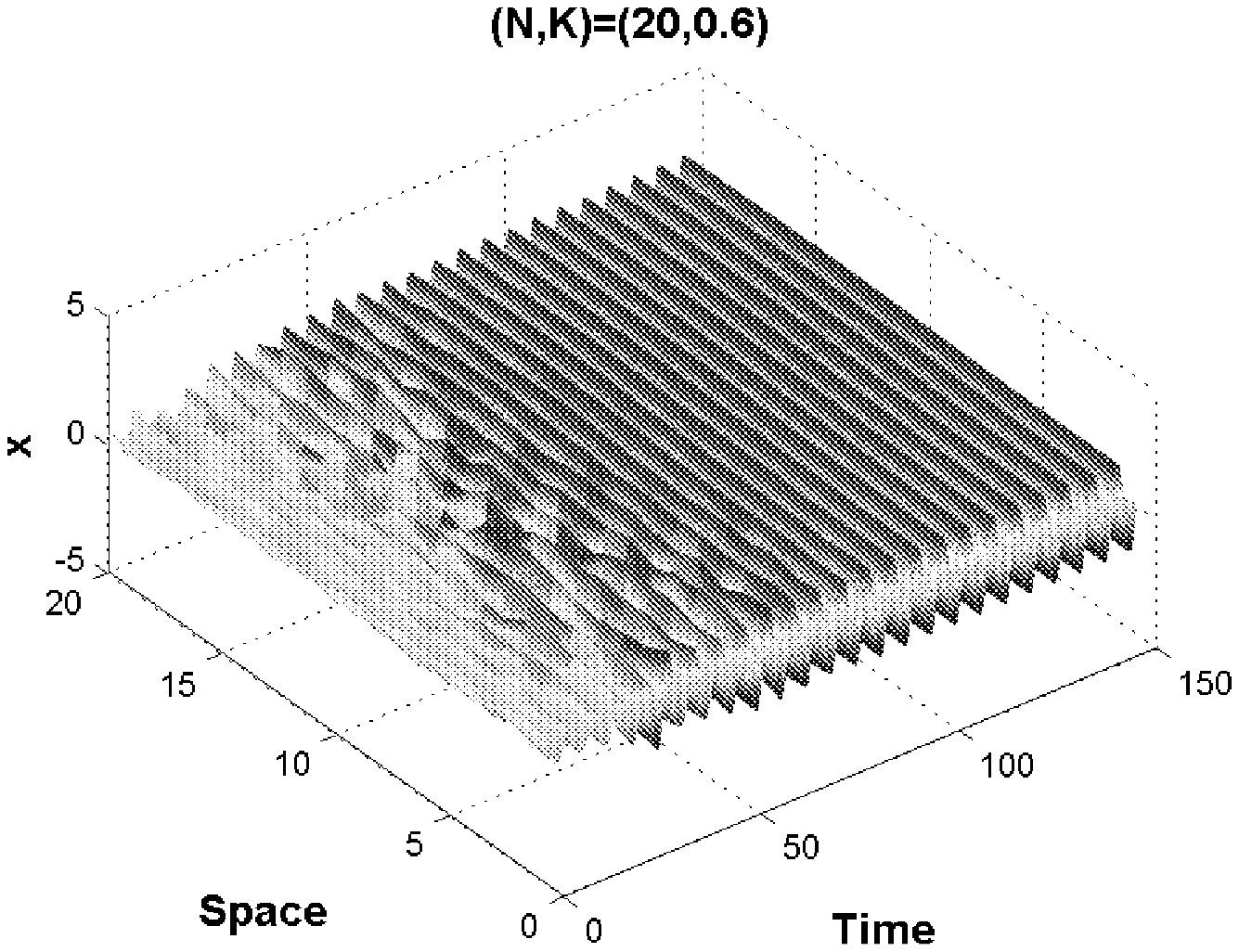}}
\end{picture}
\caption[] {    \it  Space-time-amplitude plot showing
transitions from unstable to stable synchronization in a ring of
diffusively coupled oscillators, with $\mu=\alpha=\beta=0.1$. }
\end{center}
\end{figure}

\begin{figure}[htb]
\centering
\begin{center}
\begin{picture}(150,130)
\put(-10,55)
{\includegraphics[width=7.5cm,height=5cm]{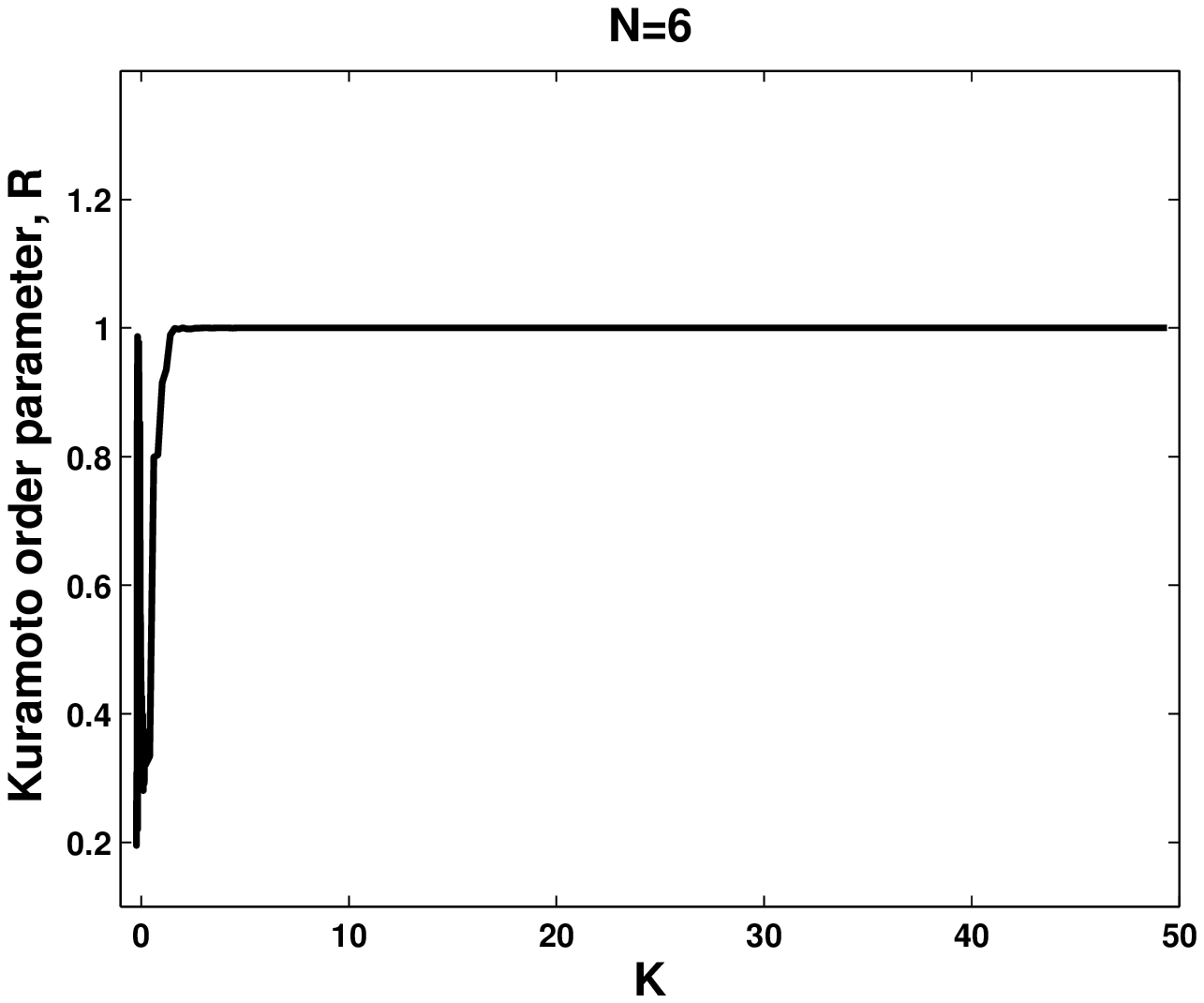}}
\put(65,55)
{\includegraphics[width=7.5cm,height=5cm]{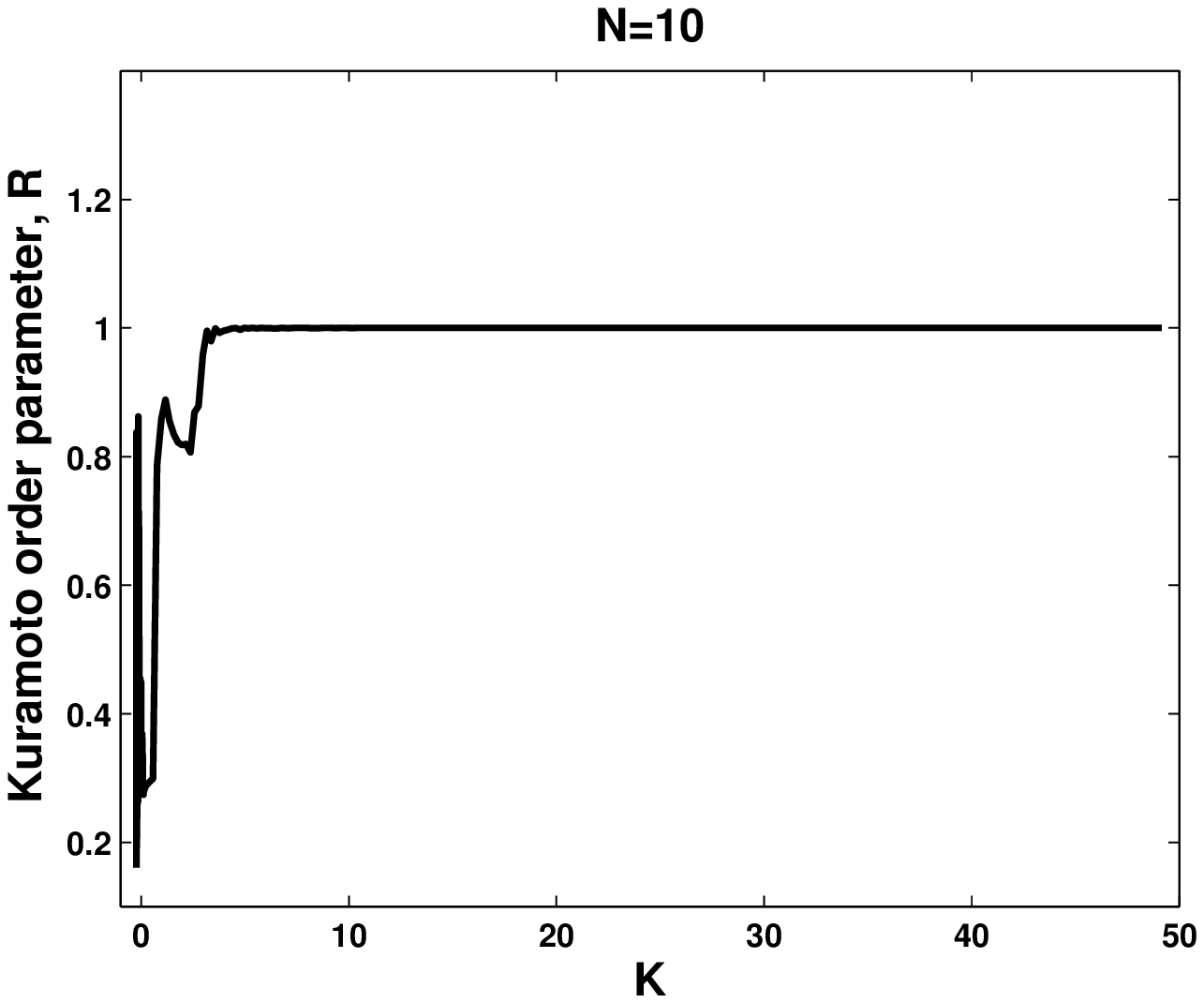}}
\put(-10,0)
{\includegraphics[width=7.5cm,height=5cm]{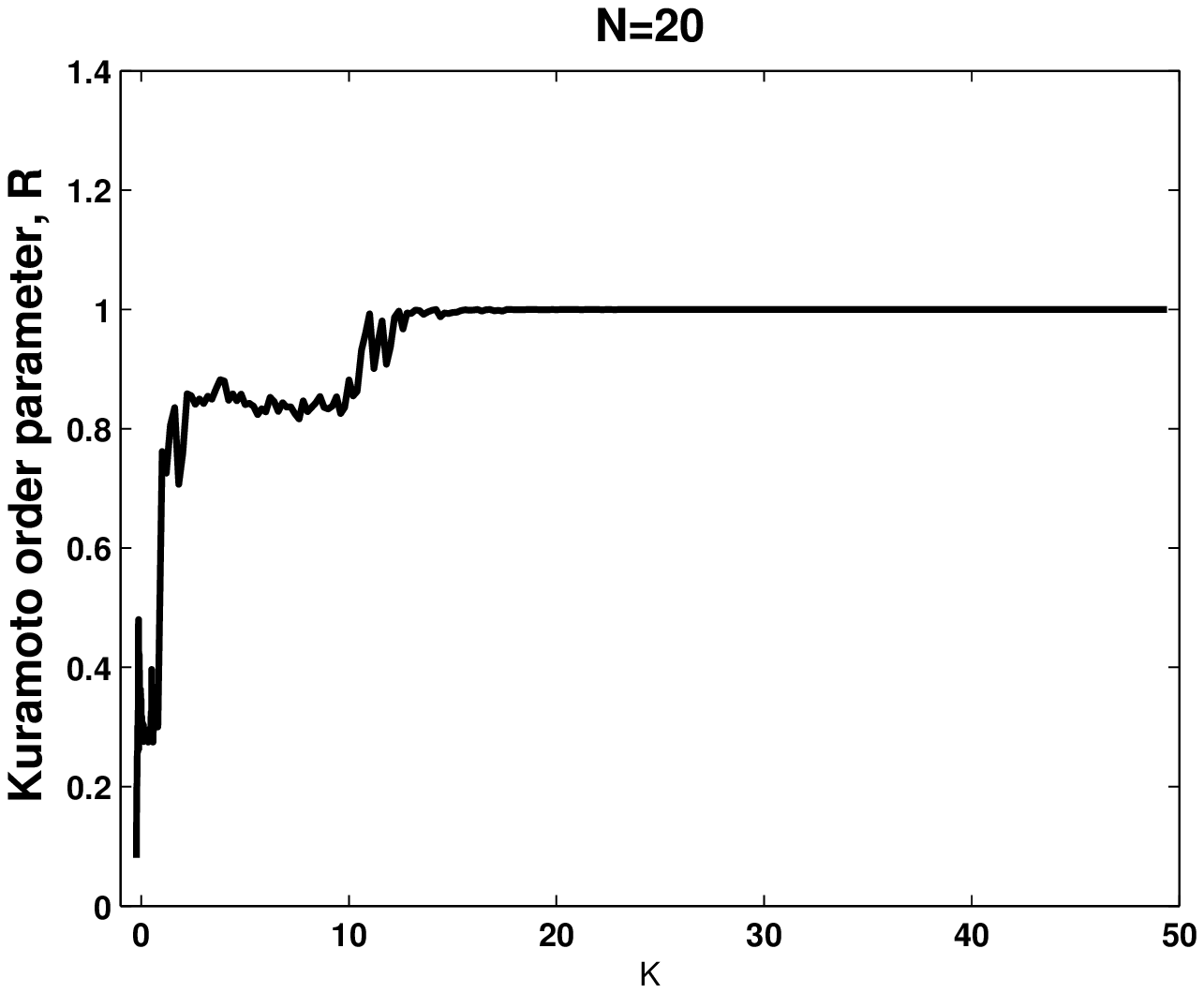}}
\put(65,0)
{\includegraphics[width=7.5cm,height=5cm]{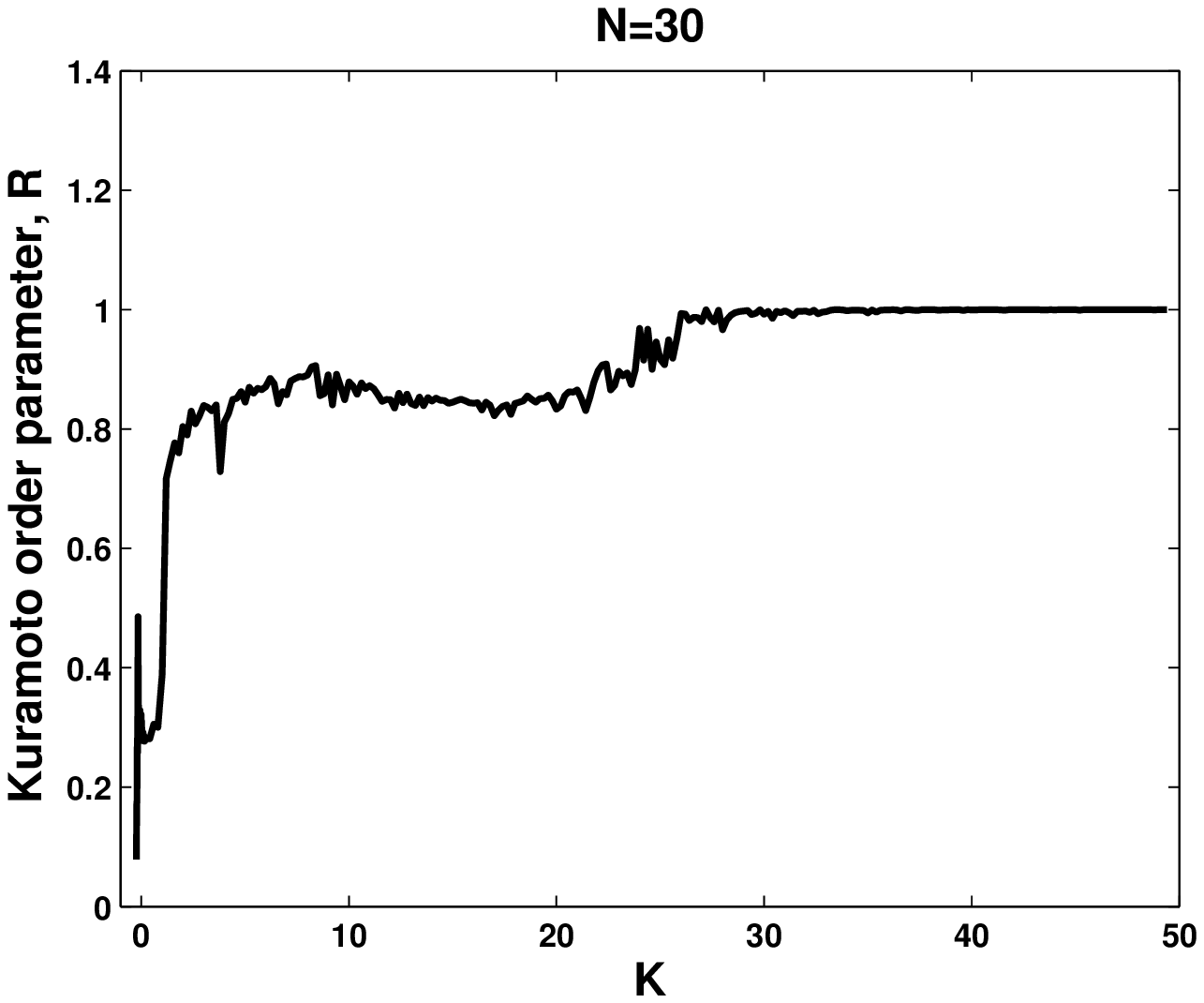}}
\end{picture}
\caption[] {    \it Variation of the Kuramoto order
parameter $R$ versus the coupling strength $K$ for a ring of
diffusively coupled non-identical oscillators with parameters $\mu
= \alpha=\beta=0.1, \Delta w_o=0.05, \kappa=2$.} 
\end{center}
\end{figure}

\begin{figure}[htb]
\centering
\begin{center}
\begin{picture}(150,130)
\put(-10,55)
{\includegraphics[width=7.5cm,height=5cm]{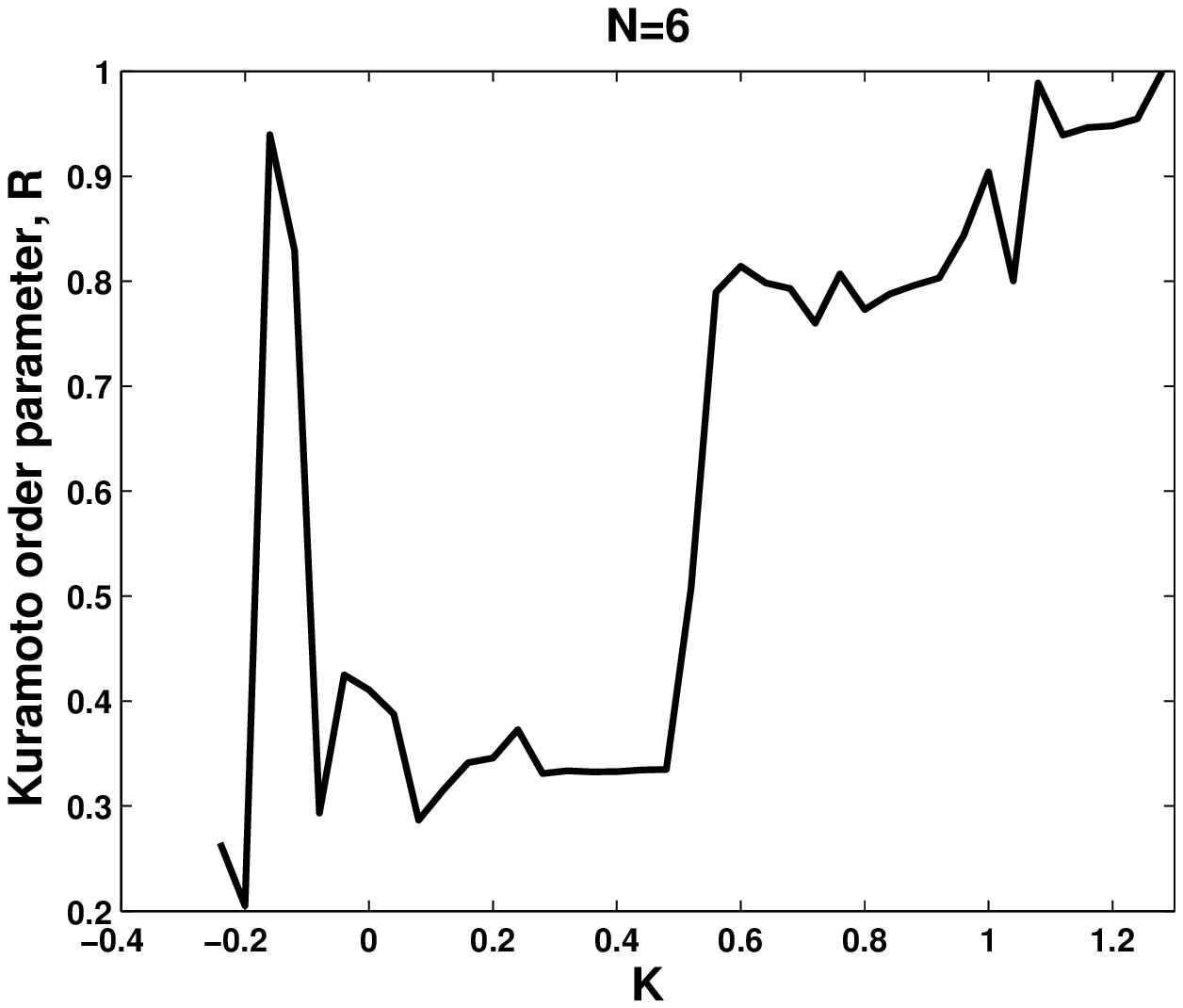}}
\put(65,55)
{\includegraphics[width=7.5cm,height=5cm]{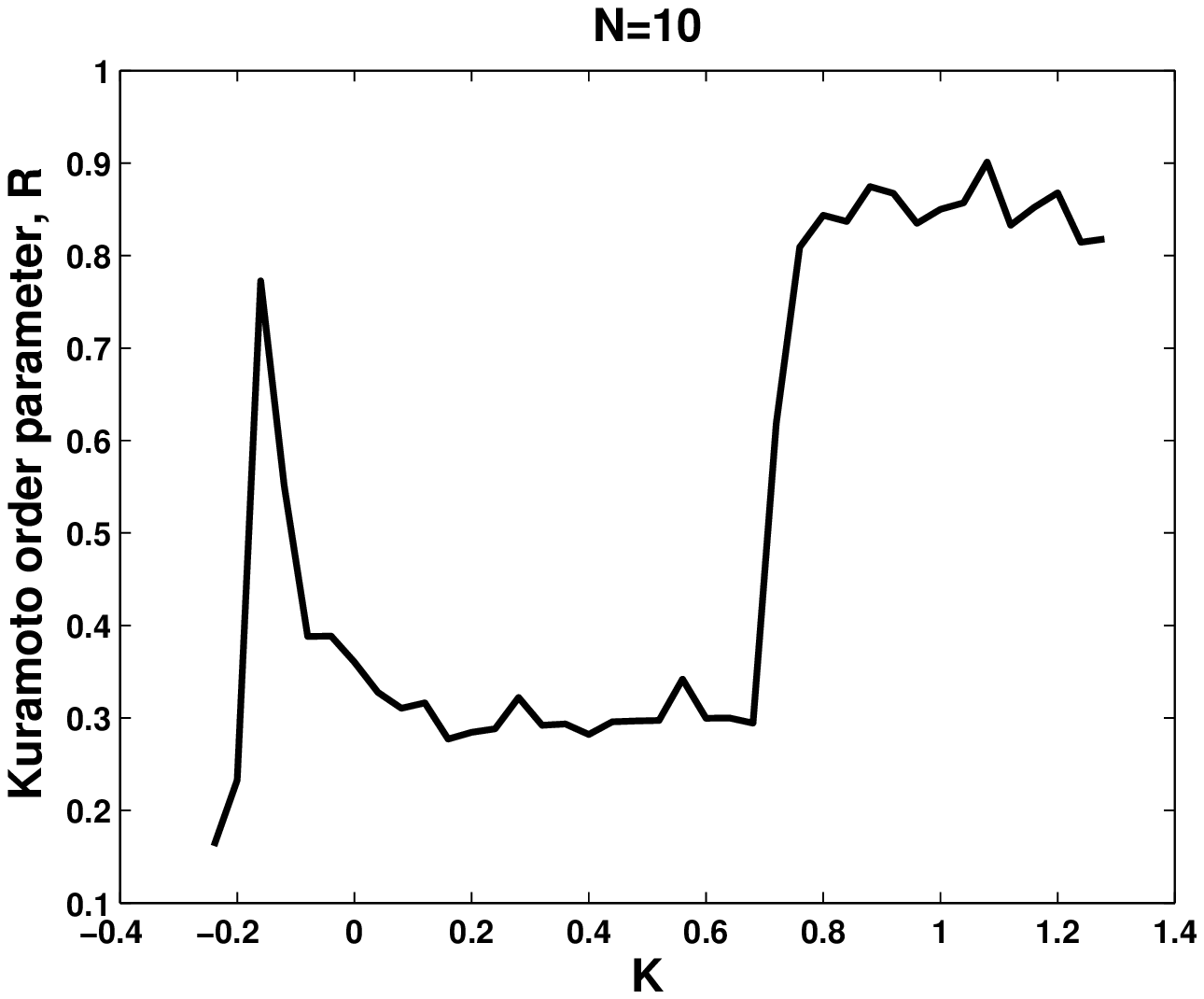}}
\put(-10,0)
{\includegraphics[width=7.5cm,height=5cm]{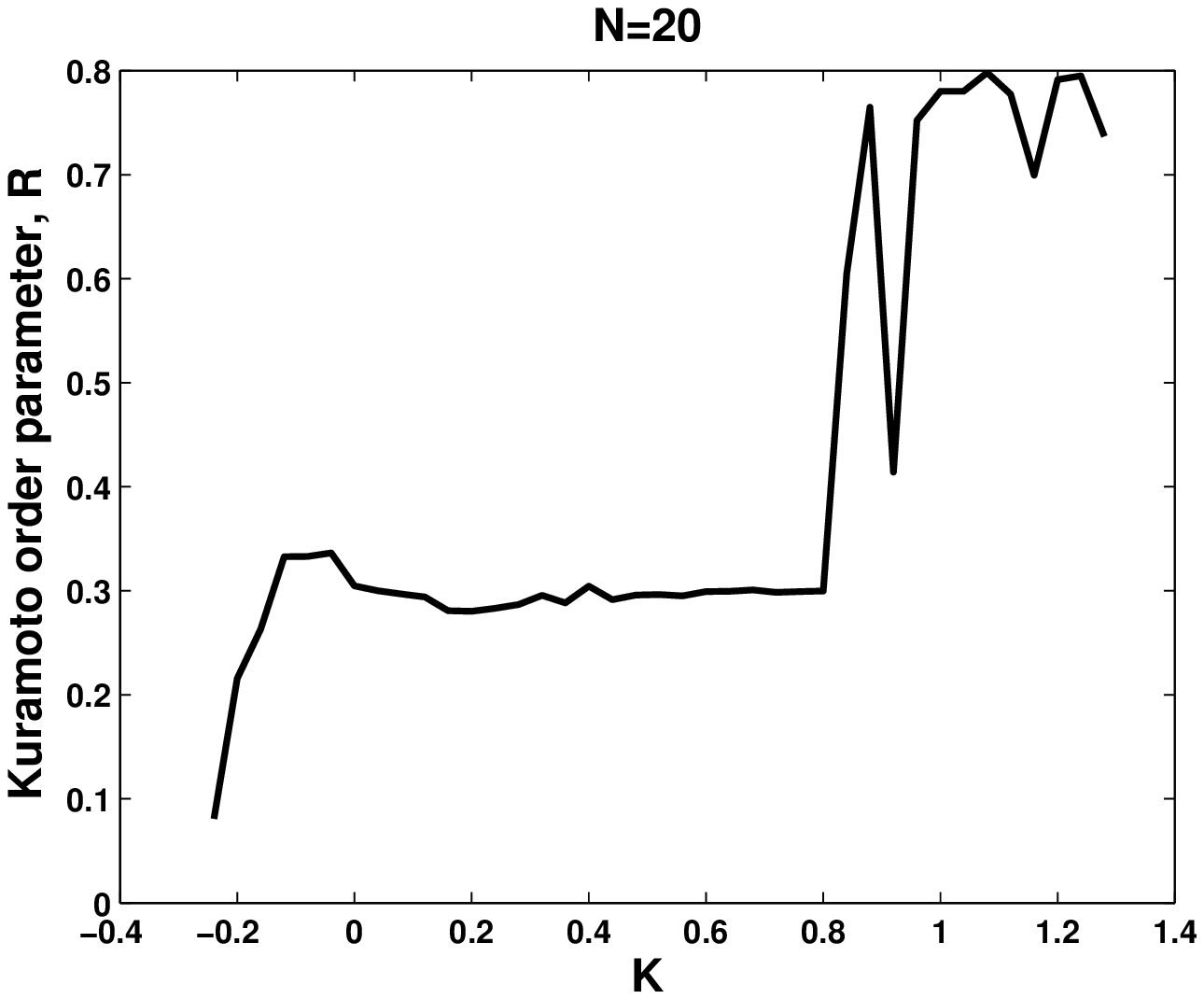}}
\put(65,0)
{\includegraphics[width=7.5cm,height=5cm]{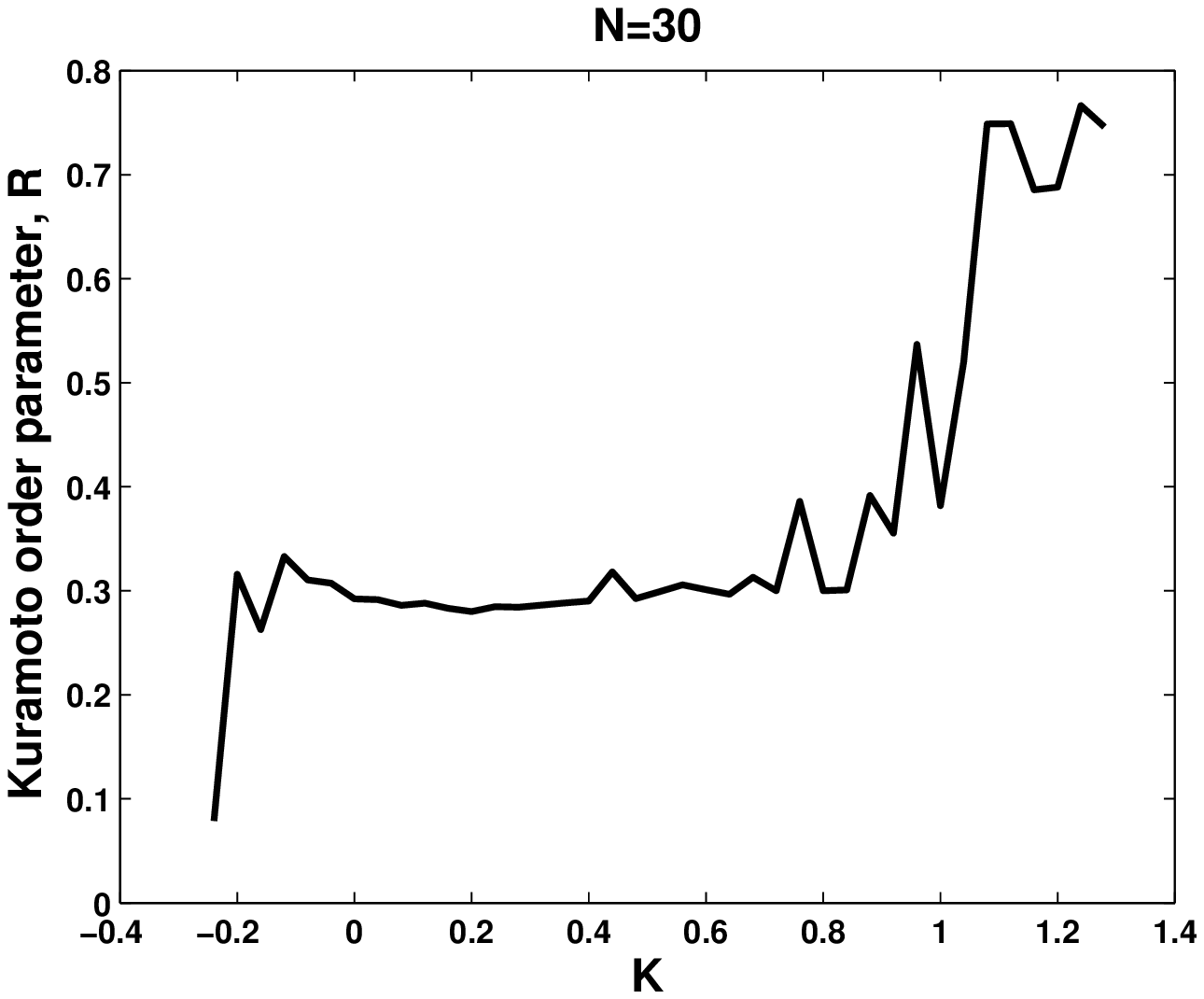}}
\end{picture}
\caption[] {    \it
Enlargement of the Kuramoto order parameter $R$ versus
the coupling strength $K$ for the same parameters as in Fig. 6
.
}
\end{center}
\end{figure}

\begin{figure}[htb]
\centering
\begin{center}
\begin{picture}(150,150)
\put(0.0,0.0)
{\includegraphics[width=12cm,height=7cm]{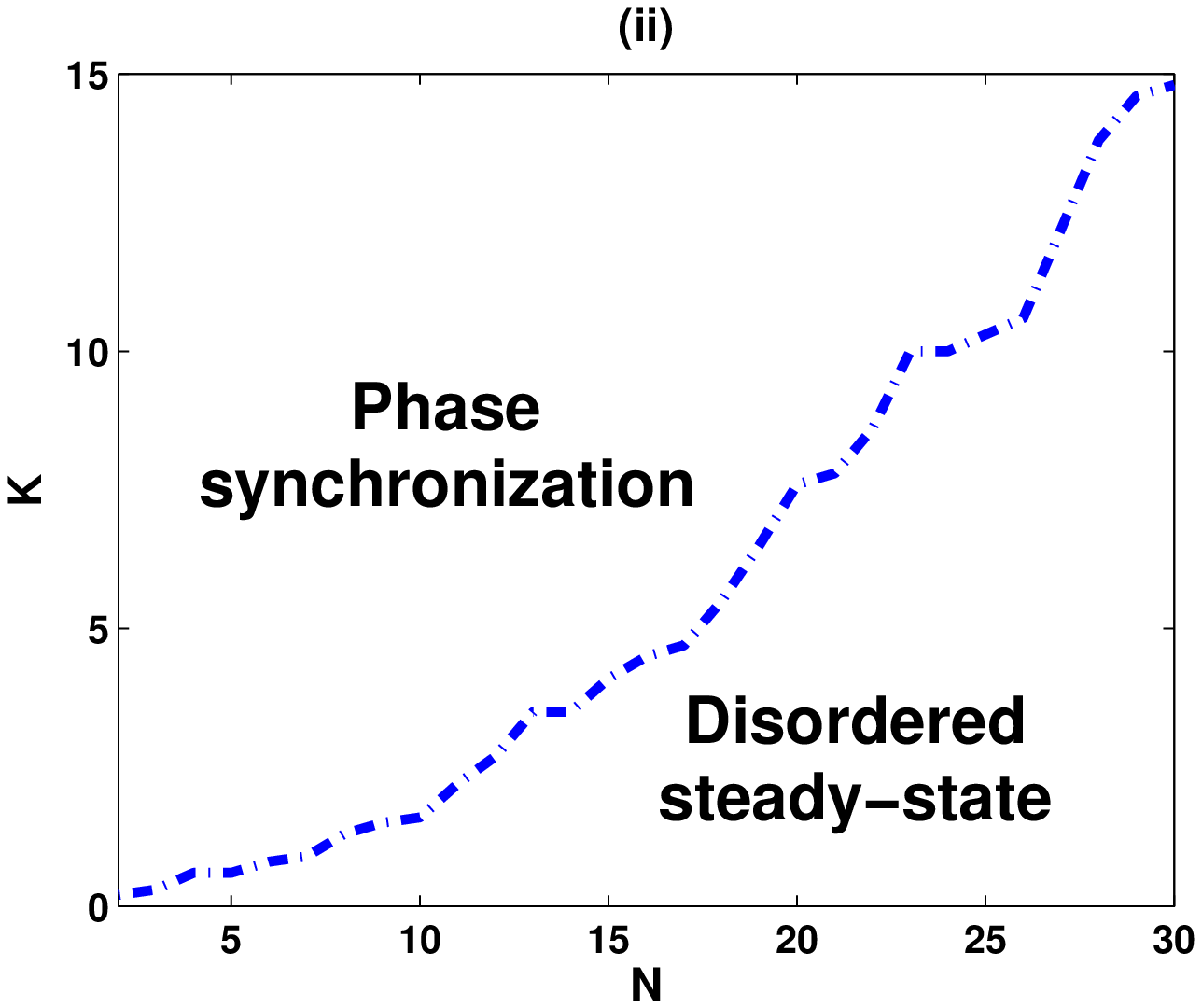}}
\put(0.0,70.0)
{\includegraphics[width=12cm,height=7cm]{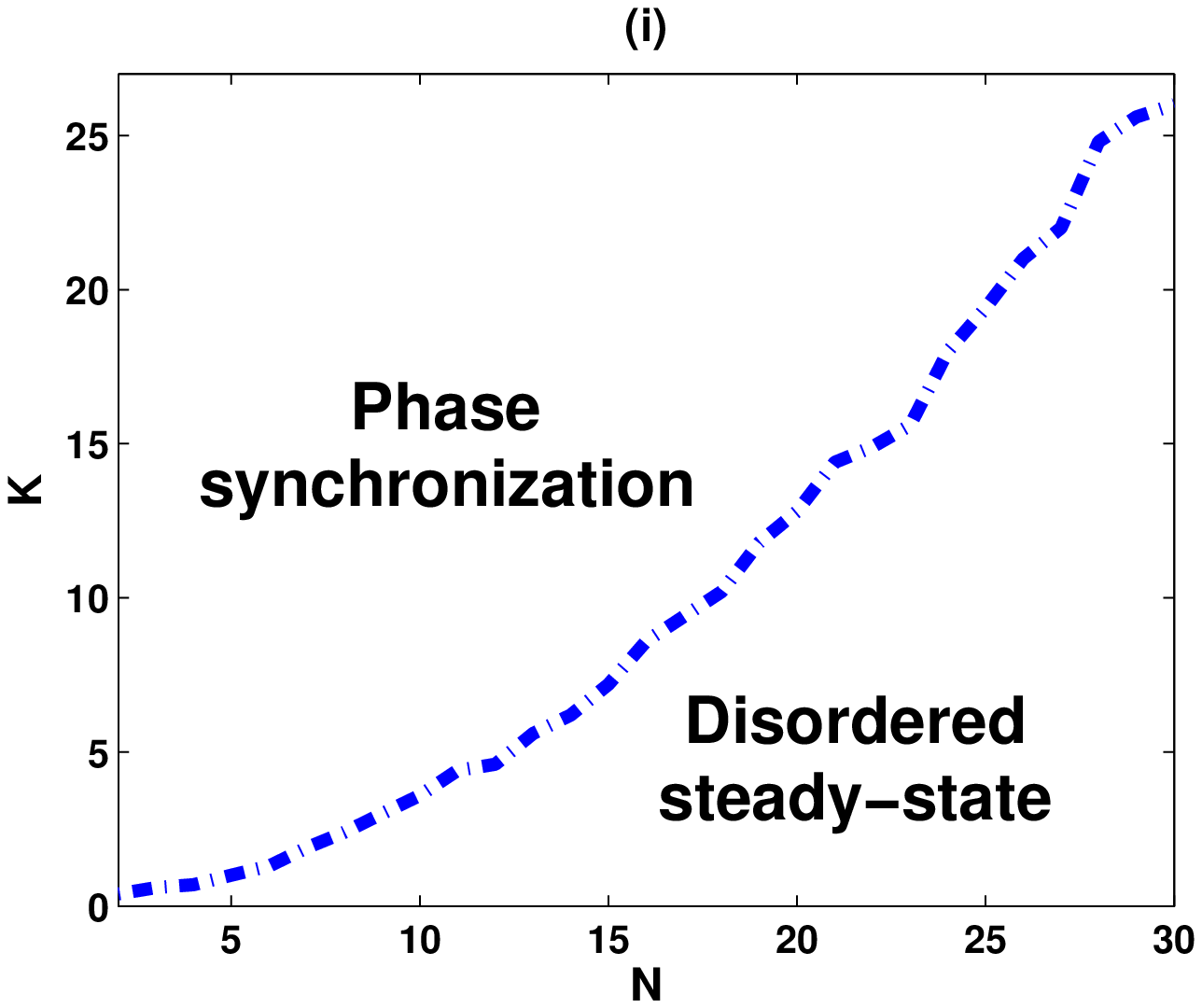}}
\end{picture}
\caption[] {    \it
Stability diagram of phase synchronization ($R>0.998$) and disordered
 steady-state ($R<0.998$) for a ring of diffusively coupled
 nonidentical oscillators. Parameters are $\mu = 0.1$,
 $\Delta w_o=0.05$, $\kappa=2$, (i): $\alpha=\beta=0.1$, (ii): $\alpha=0.114,
\beta=0.005$.}
\end{center}
\end{figure}

\begin{figure}[htb]
\centering
\begin{center}
\begin{picture}(250,150)
\put(0,70)
{\includegraphics[width=12cm,height=7cm]{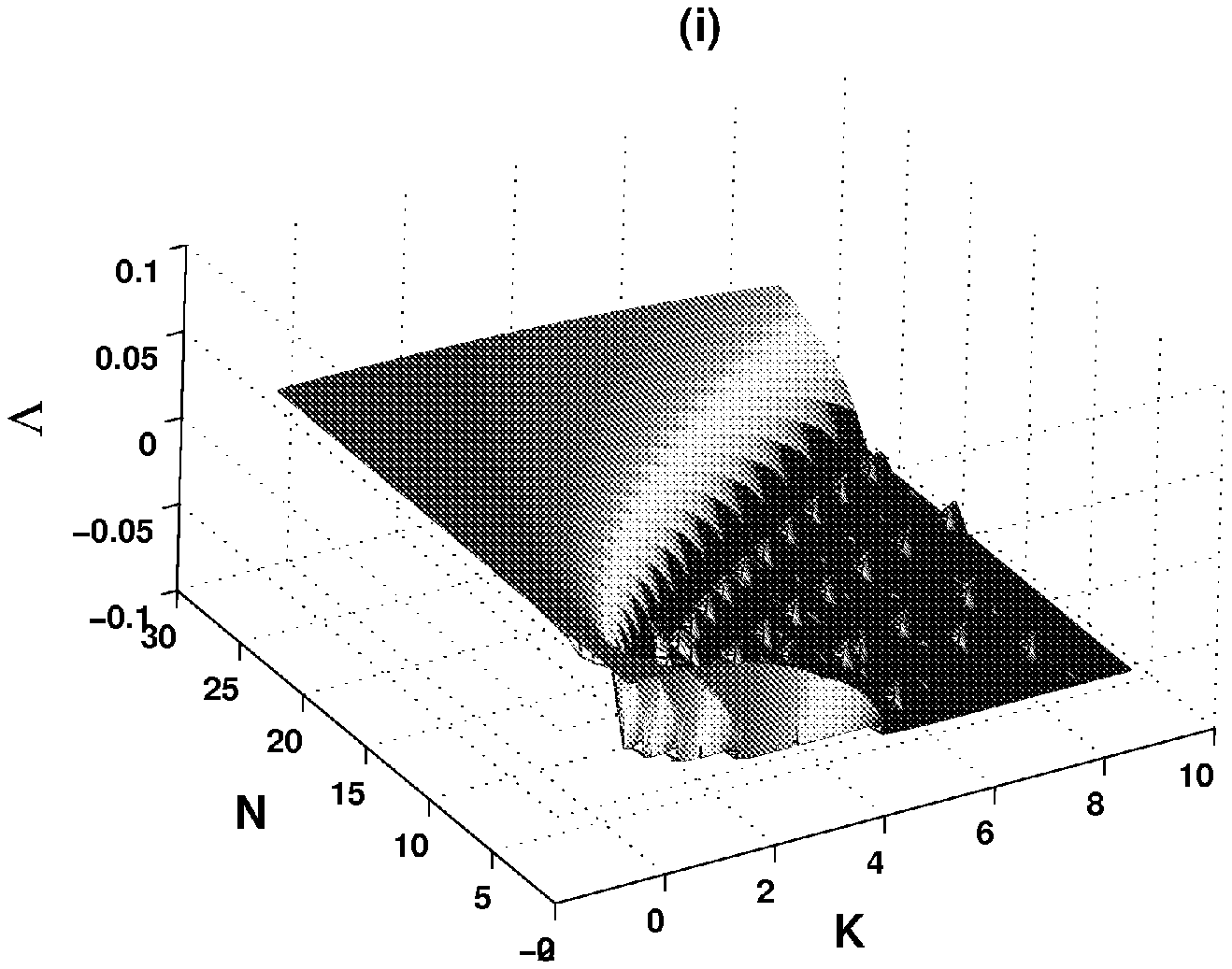}}
\put(0,0)
{\includegraphics[width=12cm,height=7cm]{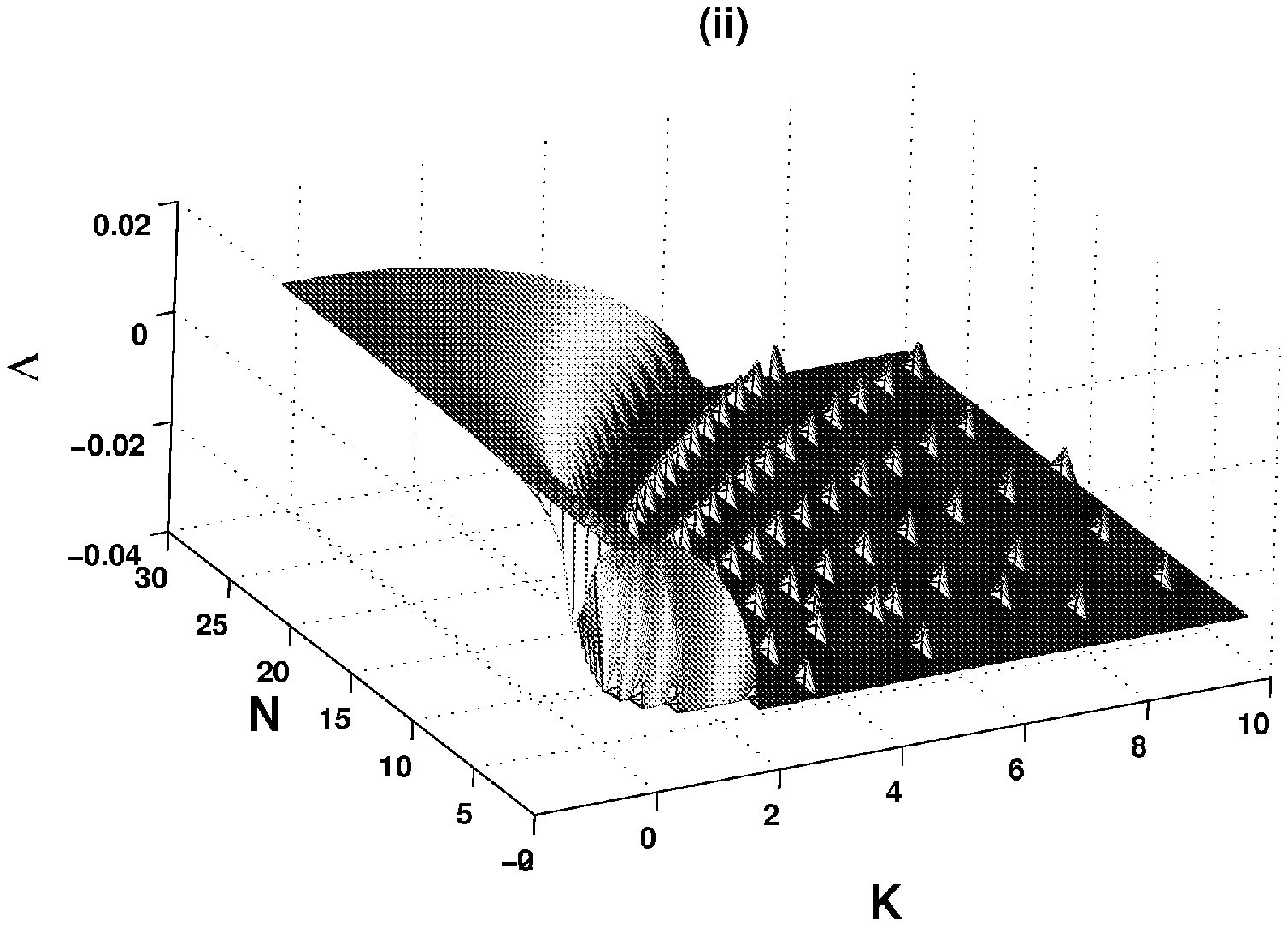}}
\end{picture}
\caption[] {    \it Variation of the Master Stability
Function (MSF) $\Lambda$ versus $K$ and $N$ for a chain of
open-ended of diffusive coupling oscillators, (i)
$\mu=\beta=\alpha=0.1$ and (ii) $\mu=0.1; \beta=0.005,
\alpha=0.114.$} 
\end{center}
\end{figure}

\begin{figure}[htb]
\centering
\begin{center}
\begin{picture}(250,150)
\put(0.0,75.0)
{\includegraphics[width=12.3cm,height=7cm]{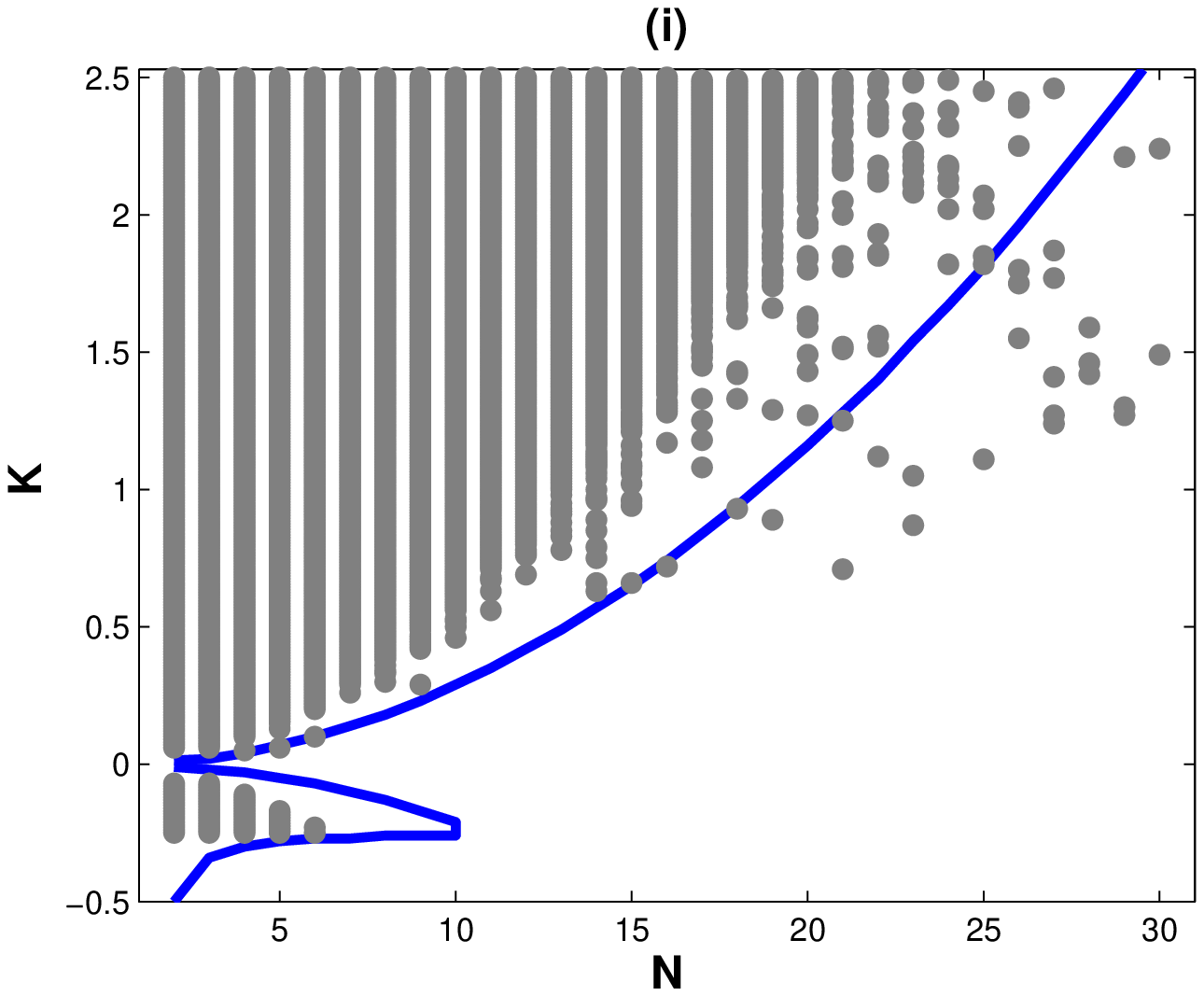}}
\put(0.0,0.0)
{\includegraphics[width=12.3cm,height=7cm]{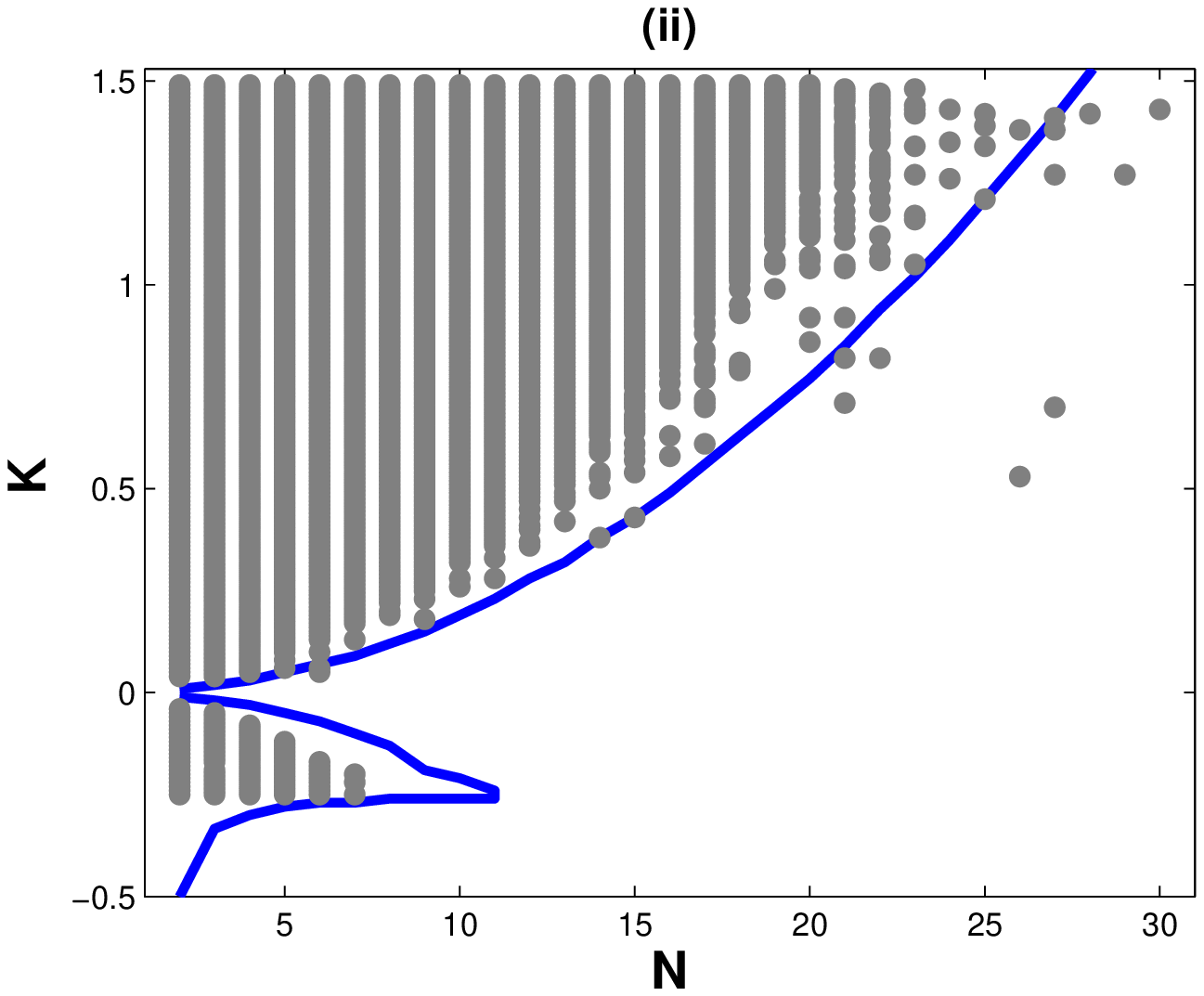}}
\end{picture}
\caption[] {    \it Stability diagram in the $(N,K)$
plane of chain of open ended diffusively coupled oscillators. The
shaded area denotes the region of stable synchronization obtained
numerically with $h=10^{-6}$ (see Eq.(\ref{eq7}) ), while the
solid line is the stability boundary obtained through the Master
Stability Function (MSF). Parameters of the system are $\mu=0.1$;
(i) $\beta=\alpha=0.1$, (ii) $\beta=0.005,\alpha=0.114$. }
\end{center}
\end{figure}

\begin{figure}[htb]
\centering
\begin{center}
\begin{picture}(150,130)
\put(-10,55)
{\includegraphics[width=7cm,height=5cm]{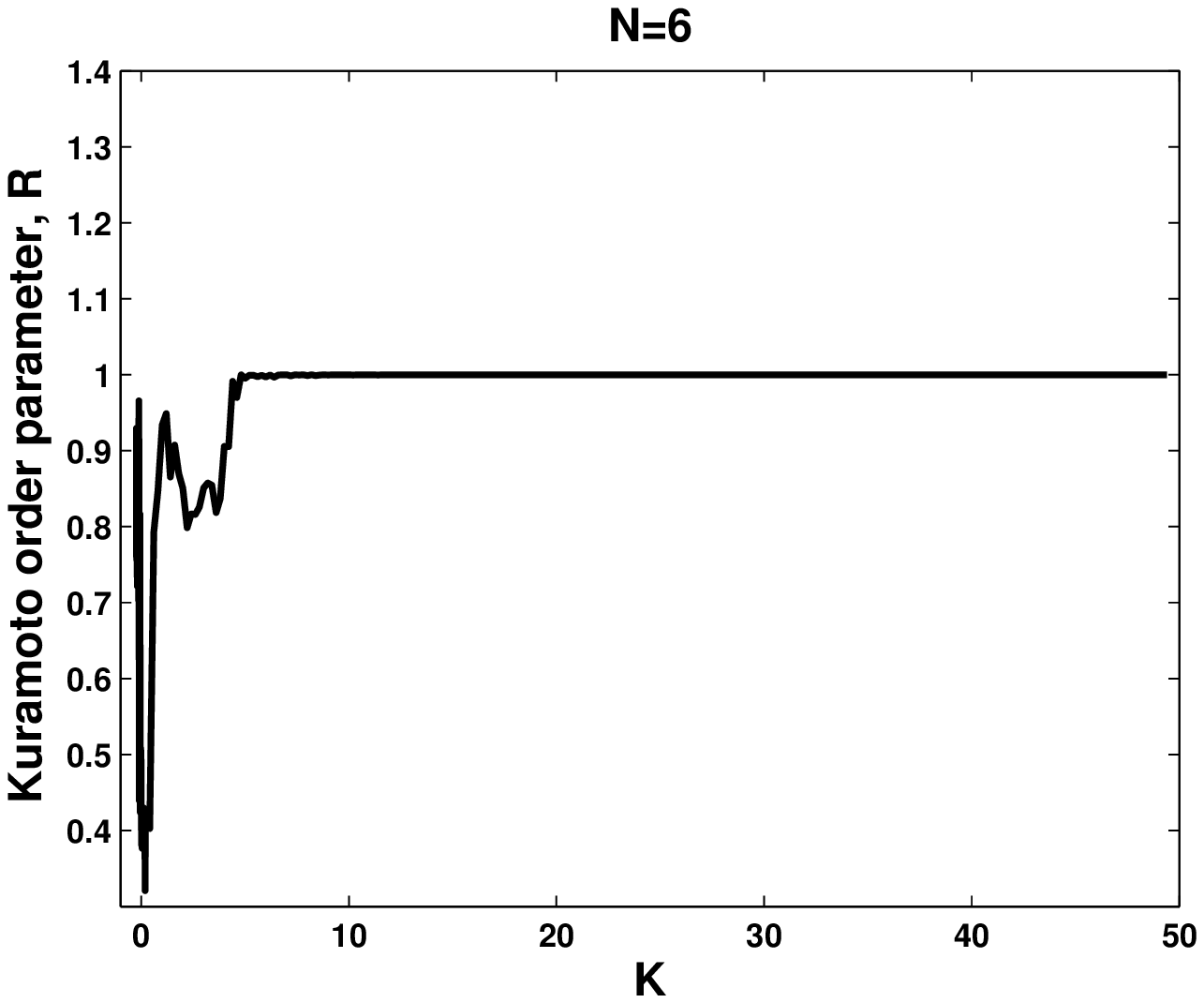}}
\put(65,55)
{\includegraphics[width=7cm,height=5cm]{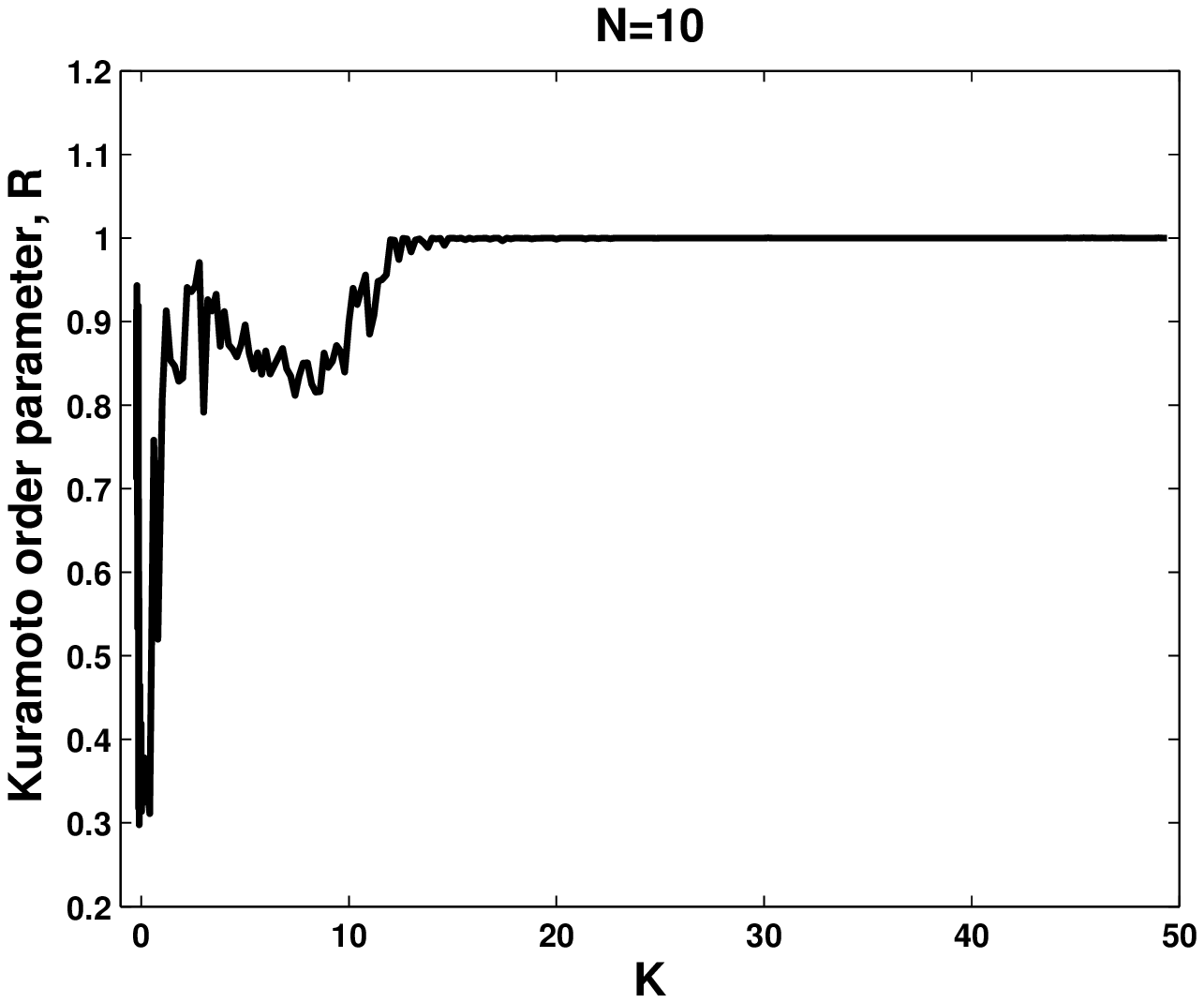}}
\put(-10,0)
{\includegraphics[width=7cm,height=5cm]{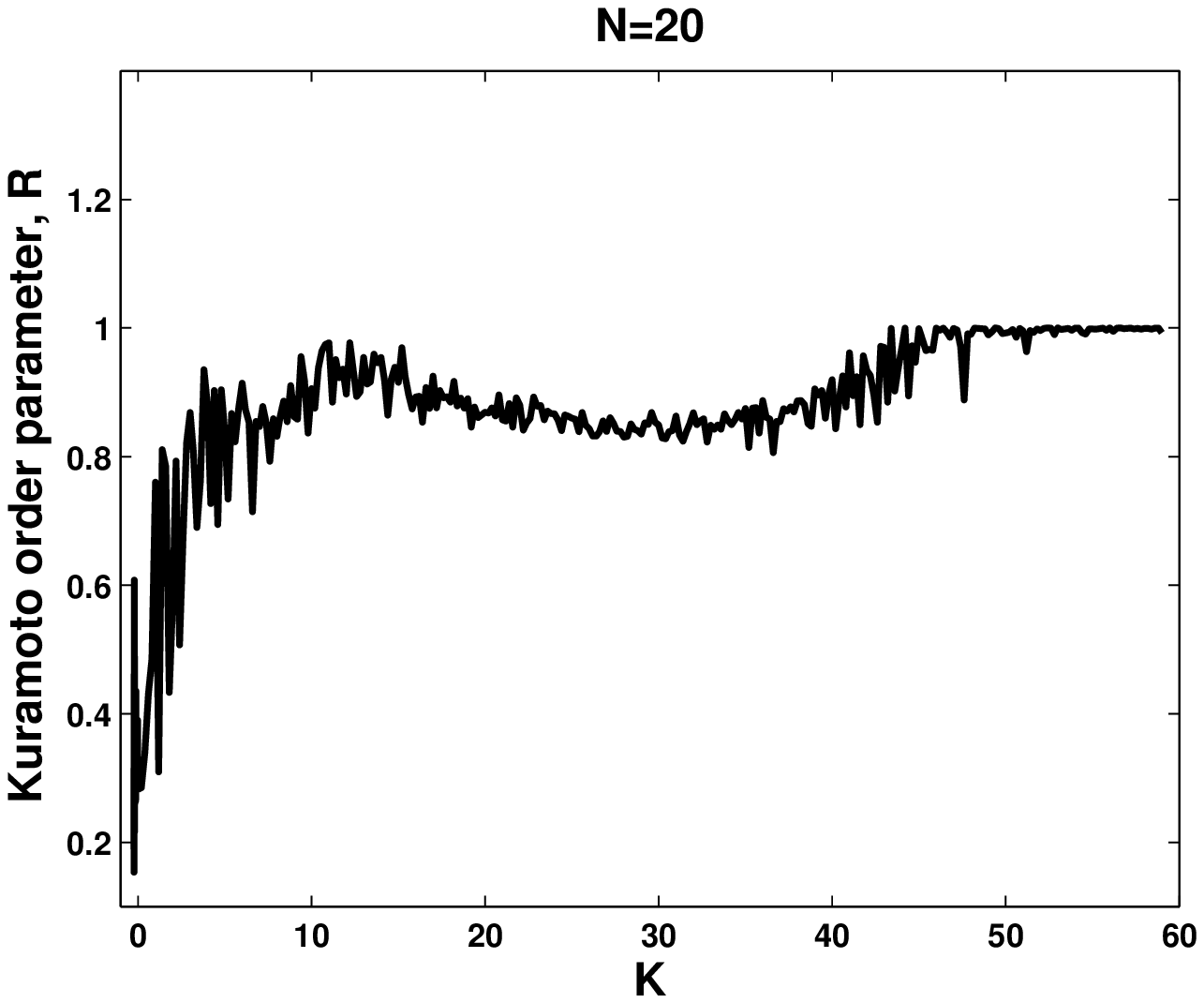}}
\put(65,0)
{\includegraphics[width=7cm,height=5cm]{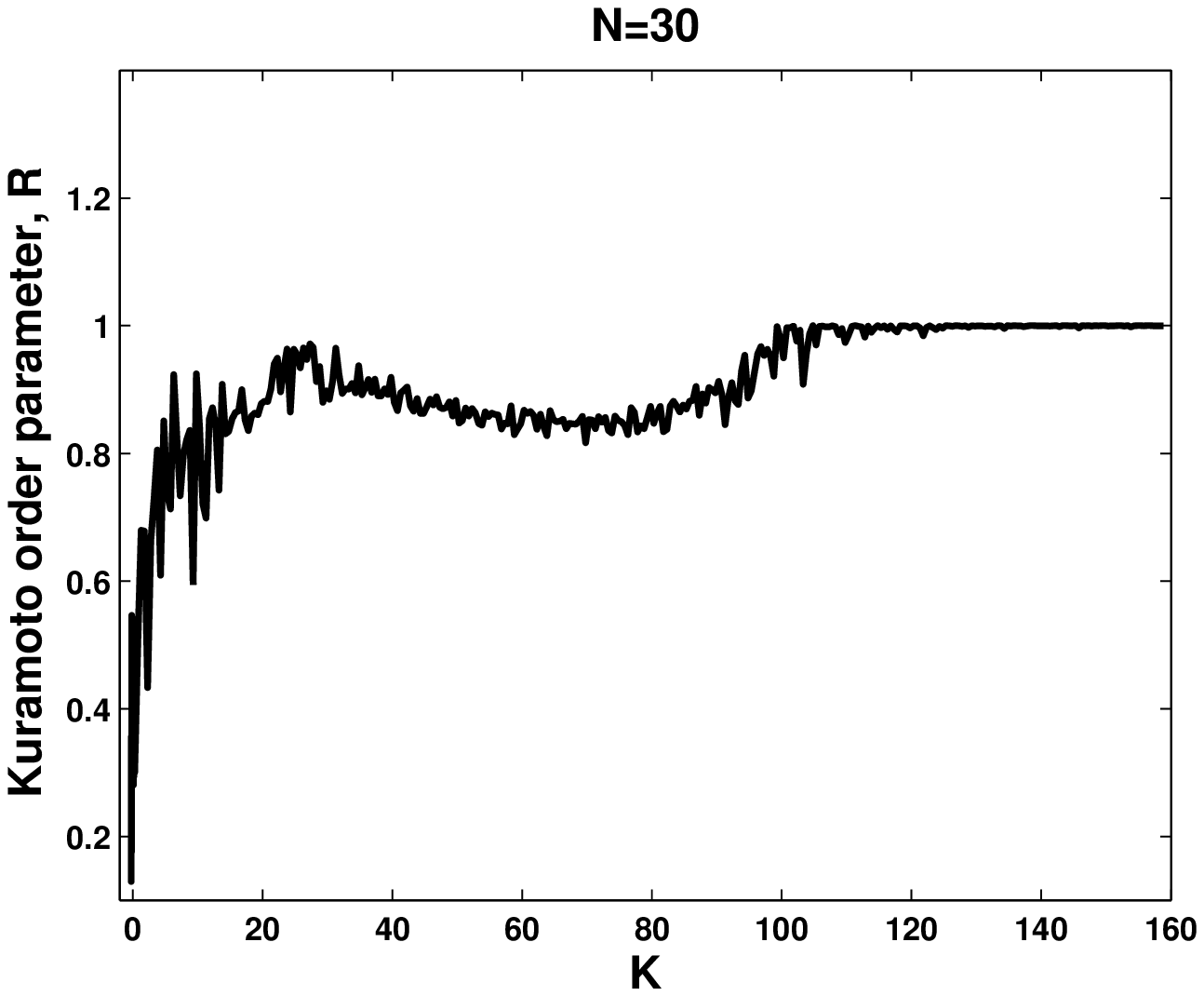}}
\end{picture}
\caption[] {    \it
Kuramoto order parameter $R$ versus
the coupling strength $K$ for a chain of open-ended diffusively
coupled non-identical oscillators with parameters
$\mu=\alpha=\beta=0.1, \Delta w_o=0.05, \kappa=2$.}
\end{center}
\end{figure}

\begin{figure}[htb]
\centering
\begin{center}
\begin{picture}(150,150)
\put(0.0,0.0)
{\includegraphics[width=12cm,height=7cm]{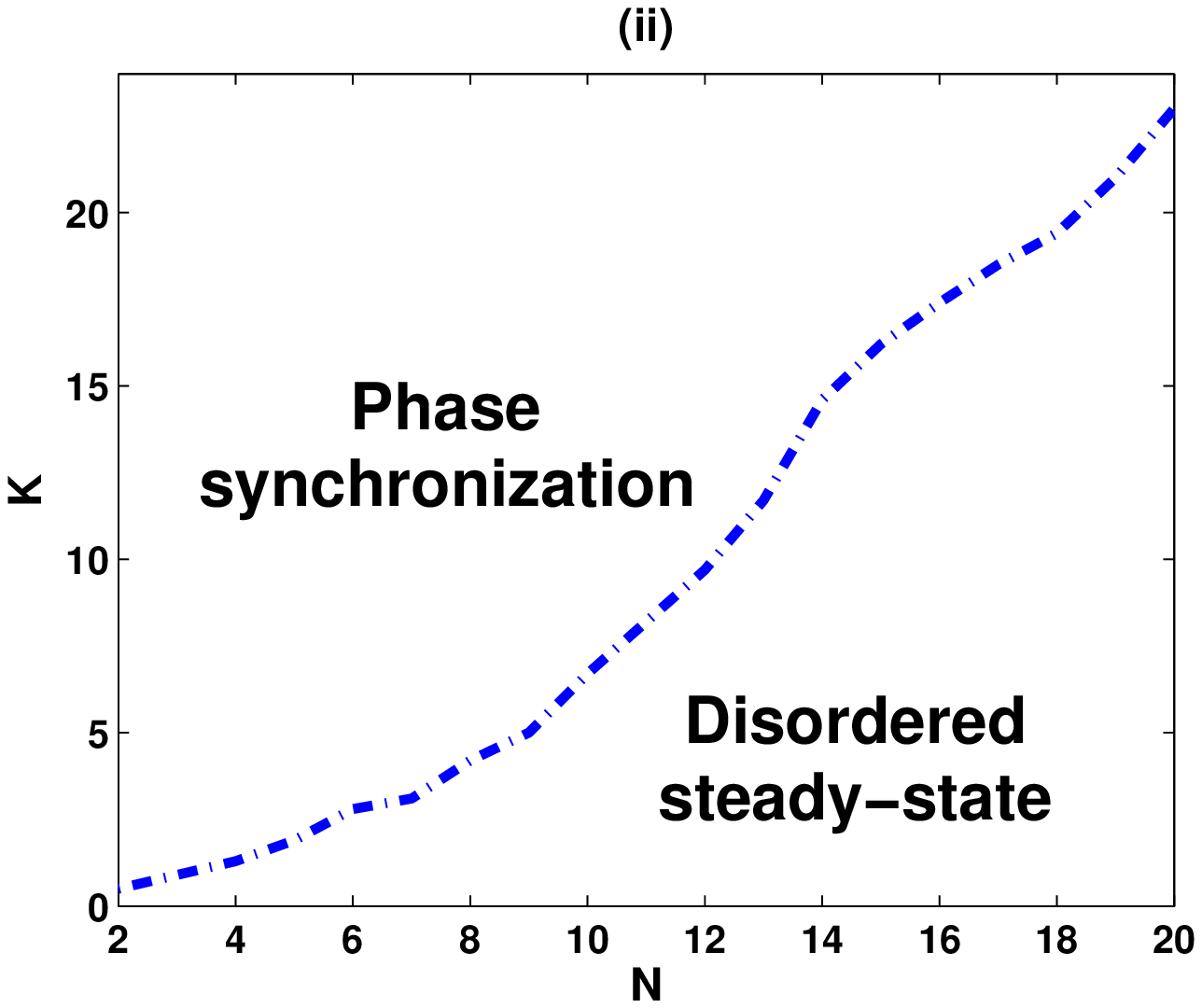}}
\put(0.0,70.0)
{\includegraphics[width=12cm,height=7cm]{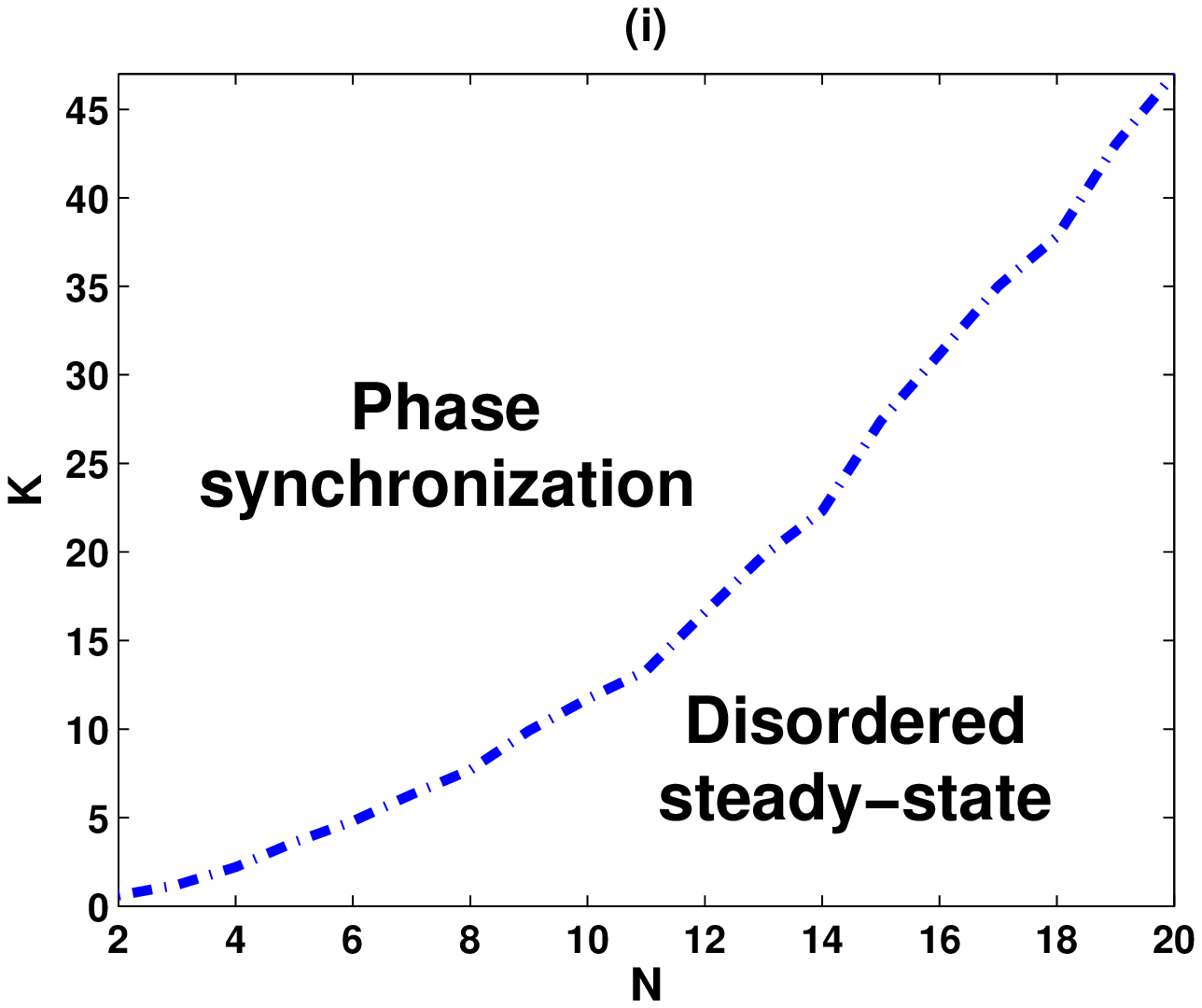}}
\end{picture}
\caption[] {    \it
Stability diagram of phase synchronization ($R>0.998$) and
disordered steady-state ($R<0.998$) for an open-ended chain of
diffusively coupled nonidentical oscillators. Parameters are
 $\mu = 0.1$, $\Delta w_o=0.05$, $\kappa=2$,
 (i): $\alpha=\beta=0.1$, (ii): $\alpha=0.114,
\beta=0.005$.}
\end{center}
\end{figure}

\end{document}